%% file: main.tex
\documentclass[9pt,shortpaper,twoside,web]{ieeecolor}
\usepackage{generic}
\usepackage{cite}
\usepackage{amsmath,amssymb,amsfonts}
\usepackage{algorithmic}
\usepackage{graphicx}
\usepackage{textcomp}
\usepackage[table,xcdraw]{xcolor}
\usepackage[acronym]{glossaries}
\usepackage{adjustbox}
\usepackage{svg}
\usepackage{makecell} 

\usepackage{acronym}
\usepackage{siunitx}
\usepackage{bm}

\include{include}

\def\BibTeX{{\rm B\kern-.05em{\sc i\kern-.025em b}\kern-.08em
    T\kern-.1667em\lower.7ex\hbox{E}\kern-.125emX}}
\makeatletter
\def\ps@IEEEtitlepagestyle{%
  \def\@oddhead{}%
  \def\@evenhead{}%
  \def\@oddfoot{%
    \hbox to\textwidth{%
      \hfil
      \raisebox{0.0\baselineskip}{
        \parbox{0.98\textwidth}{\centering\scriptsize
          \copyright\ 2025 IEEE. Personal use of this material is permitted.
          Permission from IEEE must be obtained for all other uses, in any current
          or future media, including reprinting/republishing this material for
          advertising or promotional purposes, creating new collective works, for
          resale or redistribution to servers or lists, or reuse of any copyrighted
          component of this work in other works.
        }%
      }%
      \hfil
    }%
  }%
  \def\@evenfoot{}%
}
\makeatother

\begin{document}

\acrodef{AW}{alpha waves}
\acrodef{PULP}{Parallel Ultra Low Power}
\acrodef{SoC}{System on Chip}
\acrodef{BLE}{Bluetooth Low Energy}
\acrodef{EEG}{electroencephalography}
\acrodef{EMG}{electromyography}
\acrodef{ECG}{electrocardiogram}
\acrodef{PPG}{photoplethysmogram}
\acrodef{EOG}{electrooculography}
\acrodef{MM}{Motor Movement}
\acrodef{SSVEP}{Steady State Visually Evoked Potential}
\acrodef{HMI}{human-machine interface}
\acrodef{PMIC}{Power Management Integrated Circuit}
\acrodef{IMU}{inertial measurement unit}
\acrodef{DSP}{digital signal processing}
\acrodef{NN}{neural network}
\acrodef{ULP}{ultra-low-power}
\acrodef{sEMG}{surface electromyography}
\acrodef{ML}{Machine Learning}
\acrodef{LOSO CV}{leave-one-subject-out cross-validation}
\acrodef{CV}{cross-validation}
\acrodef{BCI}{Brain-Computer Interface}
\acrodef{CCA}{canonical-correlation analysis}
\acrodef{SoA}{State-of-the-Art}
\acrodef{SNR}{signal-to-noise ratio}
\acrodef{PGA}{programmable-gain amplifier}
\acrodef{BTE}{Behind-the-Ear}
\acrodef{CNN}{convolutional neural network}
\acrodef{ITR}{information transfer rate}
\acrodef{NE16}{Neural Engine 16}
\acrodef{PTH}{plated through hole}
\acrodef{CMRR}{common-mode rejection ratio}
\acrodef{AFE}{analog frontend}
\acrodef{PCB}{printed circuit board}
\acrodef{IFCN}{International Federation of Clinical Neurophysiology}
\acrodef{NCCA}{Normalized Canonical Correlation Analysis}
\acrodef{ADC}{Analog to Digital Converter}
\acrodef{CV}{Cross Validation}
\acrodef{SoA}{State-of-the-Art}

\title{GAPses: Versatile smart glasses for comfortable and fully-dry acquisition and parallel ultra-low-power processing of EEG and EOG}

\author{Sebastian~Frey, \IEEEmembership{Graduate Student Member, IEEE}, Mattia Alberto Lucchini, Victor~Kartsch, Thorir~Mar~Ingolfsson, \IEEEmembership{Graduate Student Member, IEEE}, Andrea Helga Bernardi,  Michael Segessenmann, Jakub Osieleniec, Simone Benatti, \IEEEmembership{Member, IEEE}, Luca Benini, \IEEEmembership{Fellow, IEEE}, and Andrea Cossettini, \IEEEmembership{Member, IEEE}
\thanks{This project was supported by the Swiss National Science Foundation (Project PEDESITE) under grant agreement 193813, by the ETH-Domain Joint Initiative program (project UrbanTwin), and by the ETH Future Computing Laboratory (EFCL). }
\thanks{Sebastian Frey, Victor Kartsch, Thorir Mar Ingolfsson, Luca Benini, and Andrea Cossettini are with the Integrated Systems Laboratory, ETH Z{\"u}rich, Z{\"u}rich, Switzerland.}
\thanks{Luca Benini is also with the DEI, University of Bologna, Bologna, Italy.}
\thanks{Andrea Helga Bernardi is with the DEI, University of Bologna, Bologna, Italy.}
\thanks{Simone Benatti is with the DISMI, University of Modena and Reggio Emilia, Reggio Emilia, Italy, and with the DEI, University of Bologna, Bologna, Italy}
\thanks{Mattia Lucchini, Michael Segessenmann, Jakub Osieleniec are with D{\"a}twyler Schweiz AG, Schattdorf, Switzerland.}}

\maketitle

\thispagestyle{IEEEtitlepagestyle}  

\input{acro}
\input{Sections/01_Abstract}

\begin{IEEEkeywords}
BCI, EEG, embedded deployment, EOG, HMI, smart glasses, wearable devices
\end{IEEEkeywords}
\input{Sections/02_Introduction}

\input{Sections/03_RelatedWorks}

\input{Sections/04_SystemDesign}

\input{Sections/05_Validation}

\input{Sections/06_Applications}

\input{Sections/07_Discussion}

\input{Sections/08_Conclusion}

\section*{Acknowledgment}
\vspace{-0.1cm}
We thank L. Mei, A. Blanco Fontao, and H. Gisler (ETH Zürich) for technical support.

\bibliographystyle{IEEEtran}
\bibliography{bib}

\end{document}

%% file: include.tex
\usepackage{comment}

\usepackage{nohyperref}

\definecolor{matplotlib0}{HTML}{1f77b4}
\definecolor{matplotlib1}{HTML}{d62728}
\definecolor{matplotlib2}{HTML}{2ca02c}
\definecolor{matplotlib3}{HTML}{ff7f0e}
\definecolor{matplotlib4}{HTML}{9467bd}
\definecolor{matplotlib5}{HTML}{8c564b}
\definecolor{matplotlib6}{HTML}{e377c2}
\definecolor{matplotlib7}{HTML}{7f7f7f}
\definecolor{matplotlib8}{HTML}{bcbd22}
\definecolor{matplotlib9}{HTML}{17becf}

\usepackage{mathtools} 

\usepackage{booktabs}
\usepackage{multirow}
\usepackage{colortbl}
\usepackage{tablefootnote}
\usepackage{threeparttable}

\usepackage{tikz}
\usetikzlibrary{shapes,arrows,positioning}

\usepackage{pgfplots}
\definecolor{color0}{rgb}{0.12156862745098,0.466666666666667,0.705882352941177} 
\definecolor{color1}{rgb}{1,0.498039215686275,0.0549019607843137}
\definecolor{color2}{rgb}{0.172549019607843,0.627450980392157,0.172549019607843} 
\definecolor{color3}{rgb}{0.83921568627451,0.152941176470588,0.156862745098039} 
\definecolor{color4}{rgb}{0.580392156862745,0.403921568627451,0.741176470588235}
\definecolor{colorblue}{rgb}{0.12156862745098,0.466666666666667,0.705882352941177} 
\definecolor{colorgreen}{rgb}{0.172549019607843,0.627450980392157,0.172549019607843} 
\definecolor{colorred}{rgb}{0.83921568627451,0.152941176470588,0.156862745098039} 
\definecolor{colorblack}{rgb}{0,0,0} 
\definecolor{amber}{rgb}{1.0, 0.75, 0.0}
\definecolor{cyan(process)}{rgb}{0.0, 0.72, 0.92}
\definecolor{colororange}{rgb}{1,0.56,0} 
\definecolor{colorbrown}{rgb}{0.62, 0.42, 0.21} 
\definecolor{colorpurple}{rgb}{0.33, 0.033, 0.51}
\usepgfplotslibrary{fillbetween}
\usepgfplotslibrary{colormaps}
\pgfplotsset{compat=1.16}

\pgfplotscreateplotcyclelist{matplotlib}{
  {matplotlib0},
  {matplotlib1},
  {matplotlib2},
  {matplotlib3},
  {matplotlib4},
  {matplotlib5},
  {matplotlib6},
  {matplotlib7},
  {matplotlib8},
  {matplotlib9}
}
\pgfplotsset{every axis/.append style={
    cycle list name=matplotlib,
}}
\pgfdeclareplotmark{mystar}{
    \node[star,star point ratio=2.25,minimum size=6pt,
          inner sep=0pt,draw=black,solid,fill=red] {};
}

\usepackage{listings}
\usepgfplotslibrary{groupplots}

\definecolor{code_default}{HTML}{000000}
\definecolor{code_keyword}{HTML}{AC4142}
\definecolor{code_identifier}{HTML}{D28445}

\lstdefinelanguage{RISCV}{
  sensitive=false,
  morecomment=[l]{//},
  alsoletter={.},
  morekeywords=[1]{
    lp.setup, mv, lw, p.lw, sw, p.sw, pv.sdotsp.b, pv.shuffle2.b, p.subNR, p.addNR
  },
  morekeywords=[2]{
    zero, ra, sp, gp, tp, t0, t1, t2, t3, t4, t5, t6, s0, s1, a0, a1, a2, a3, a4, a5, a6, a7, a8, a9, a10, a11,
  },
  morestring=[b]",
  morestring=[b]',
}[strings, comments, keywords]

\lstdefinestyle{RISCV_STYLE}{
  language=RISCV,
  numbers=none,
  basicstyle=\scriptsize\ttfamily\color{code_default},
  keywordstyle=[1]\color{matplotlib0},
  keywordstyle=[2]\color{matplotlib1},
  float,
  captionpos=b,
  belowskip=-0.5cm
}



%% file: acro.tex
\newacronym{ofa}{OFA}{Once-For-All}
\newacronym{simd}{SIMD}{Single Instruction, Multiple Data}
\newacronym{elu}{ELU}{Exponential Linear Unit}
\newacronym{relu}{ReLU}{Rectified Linear Unit}
\newacronym{rpr}{RPR}{Random Partition Relaxation}
\newacronym{mac}{MAC}{Multiply Accumulate}
\newacronym{dma}{DMA}{Direct Memory Access}
\newacronym{bmi}{BMI}{Brain--Machine Interface}
\newacronym{bci}{BCI}{Brain--Computer Interface}
\newacronym{smr}{SMR}{Sensory Motor Rythms}
\newacronym{eeg}{EEG}{Electroencephalography}
\newacronym{svm}{SVM}{Support Vector Machine}
\newacronym{svd}{SVD}{Singular Value Decomposition}
\newacronym{evd}{EVD}{Eigendecomposition}
\newacronym{iir}{IIR}{Infinite Impulse Response}
\newacronym{fir}{FIR}{Finite Impulse Response}
\newacronym{fc}{FC}{Fabric Controller}
\newacronym{nn}{NN}{Neural Network}
\newacronym{mrc}{MRC}{Multiscale Riemannian Classifier}
\newacronym{flop}{FLOP}{Floating Point Operation}
\newacronym{sos}{SOS}{Second-Order Section}
\newacronym{ipc}{IPC}{Instructions per Cycle}
\newacronym{tcdm}{TCDM}{Tightly Coupled Data Memory}
\newacronym{fpu}{FPU}{Floating Point Unit}
\newacronym{fma}{FMA}{Fused Multiply Add}
\newacronym{alu}{ALU}{Arithmetic Logic Unit}
\newacronym{dsp}{DSP}{Digital Signal Processing}
\newacronym{gpu}{GPU}{Graphics Processing Unit}
\newacronym{soc}{SoC}{System-on-Chip}
\newacronym{mi}{MI}{Motor-Imagery}
\newacronym{csp}{CSP}{Commmon Spatial Patterns}
\newacronym{fbcsp}{FBCSP}{Filter-Bank \acrlong{csp}}
\newacronym{pulp}{PULP}{parallel ultra-low power}
\newacronym{soa}{SoA}{state-of-the-art}
\newacronym{bn}{BN}{Batch Normalization}
\newacronym{isa}{ISA}{Instruction Set Architecture}
\newacronym{ecg}{ECG}{Electrocardiogram}
\newacronym{mcu}{MCU}{microcontroller}
\newacronym{rnn}{RNN}{recurrent neural network}
\newacronym{cnn}{CNN}{convolutional neural network}
\newacronym{tcn}{TCN}{temporal convolutional network}
\newacronym{emu}{EMU}{epilepsy monitoring unit}
\newacronym{ml}{ML}{Machine Learning}
\newacronym{dl}{DL}{Deep Learning}
\newacronym{ai}{AI}{Artificial Intelligence}
\newacronym{tcp}{TCP}{Temporal Central Parasagittal}
\newacronym{loocv}{LOOCV}{Leave-One-Out Cross-Validation}
\newacronym{wfcv}{WFCV}{Walk-Forward Cross-Validation}
\newacronym{rwcv}{RWCV}{Rolling Window Cross-Validation}
\newacronym{iot}{IoT}{Internet of Things}
\newacronym{auc}{AUC}{Area Under the Receiver Operator Characteristic}
\newacronym{dwt}{DWT}{Discrete Wavelet Transform}
\newacronym{fft}{FFT}{Fast Fourier Transform}
\newacronym{tpot}{TPOT}{Tree-based Pipeline Optimization Tool}

\newacronym{tuar}{TUAR}{Temple University Artifact Corpus}
\newacronym{tuev}{TUEV}{Temple University Event Corpus}

\newacronym{bss}{BSS}{Blind Source Separation}
\newacronym{ica}{ICA}{Independent Component Analysis}
\newacronym{ic}{ICs}{Independent Components}
\newacronym{asr}{ASR}{Artifact Subspace Reconstruction}
\newacronym{pca}{PCA}{Principal Component Analysis}
\newacronym{gap}{GAP}{Global Average Pooling}
\newacronym{fcn}{FCN}{Fully Connected Networks}
\newacronym{mlp}{MLP}{Multi-Layer Perceptron}
\newacronym{nas}{NAS}{Neural Architectural Search}
\newacronym{fph}{FP/h}{False Positives per Hour}
\newacronym{bvp}{BVP}{Blood volume Pulse}
\newacronym{eda}{EDA}{Electrodermal Activity}
\newacronym{acc}{ACC}{Accelerometer}
\newacronym{cae}{CAE}{Convolutional Autoencoder}
\newacronym{sswce}{SSWCE}{Sensitivity-Specificity Weighted Cross-Entropy}
\newacronym{ce}{CE}{Cross-Entropy}
\newacronym{ppg}{PPG}{Plethysmography}


%% file: Sections/01_Abstract.tex
\begin{abstract}
Recent advancements in head-mounted wearable technology are revolutionizing the field of biopotential measurement, but the integration of these technologies into practical, user-friendly devices remains challenging due to issues with design intrusiveness, comfort, reliability, and data privacy. 
To address these challenges, this paper presents \textsc{GAPses}, a novel smart glasses platform designed for unobtrusive, comfortable, and secure acquisition and processing of electroencephalography (EEG) and electrooculography (EOG) signals.
We introduce a direct electrode-electronics interface within a sleek frame design, with custom fully dry soft electrodes to enhance comfort for long wear. The fully assembled glasses, including electronics, weigh 40~g and have a compact size of 160~mm x 145~mm. An integrated parallel ultra-low-power RISC-V processor (GAP9, Greenwaves Technologies) processes data at the edge, thereby eliminating the need for continuous data streaming through a wireless link, enhancing privacy, and increasing system reliability in adverse channel conditions. We demonstrate the broad applicability of the designed prototype through validation in a number of EEG-based interaction tasks, including alpha waves, steady-state visual evoked potential analysis, and motor movement classification. Furthermore, we demonstrate an EEG-based biometric subject recognition task, where we reach a sensitivity and specificity of 98.87\% and 99.86\% respectively, with only 8 EEG channels and an energy consumption per inference on the edge as low as 121~\bm{$\mu$}J.
Moreover, in an EOG-based eye movement classification task, we reach an accuracy of 96.68\% on 11 classes, resulting in an information transfer rate of 94.78~bit/min, which can be further increased to 161.43~bit/min by reducing the accuracy to 81.43\%. The deployed implementation has an energy consumption of 40~\bm{$\mu$}J per inference and a total system power of only 12.4~mW, of which only 1.61\% is used for classification, allowing for continuous operation of more than 22~h with a small 75~mAh battery.

\end{abstract}

%% file: Sections/02_Introduction.tex
\section{Introduction}

Biopotential measurement is currently undergoing a paradigm shift towards wearable technology, impacting the way we monitor our physiological functions and interact with our surroundings by enabling continuous, long-term monitoring \cite{Dias2018_wearable_health_systems_and_technologies}. 
Continuous monitoring has the potential to deepen our understanding of human physiological responses in various contexts, advancing assistive technologies and human-machine interaction beyond research laboratory settings \cite{Yilmaz2010_wireless_sensors_vital_signs}.

Various body locations are suitable for measuring biosignals, but the human head offers unparalleled opportunities for monitoring health, drowsiness, and cognitive states, thanks to the proximity to the brain and multiple sensory organs (e.g., the eyes, nose, and ears) \cite{hussain2020healthsos}. For example, head-mounted wearables such as the cEEGrid \cite{openbci} or the Emotiv Insight and EPOC+ \cite{emotiv} demonstrated advanced capabilities in neural signals monitoring, offering insights into brain activity for cognitive studies, mental health monitoring, and \acp{BCI} \cite{Casson2019_wearable_EEG}. 

However, current research on wearable biopotential measurement devices faces significant challenges that impede widespread user acceptance and practical usage in unconstrained settings.
The main limitations of existing head-mounted biopotential measurement systems include: (1) user acceptance hindered by obtrusive designs and unconventional aesthetics of the devices \cite{Park2020_user_acceptance_wearables}; (2) achieving wearability and comfort remains difficult without the development of custom, soft electrodes that can conform to the individual's body without causing irritation or discomfort \cite{Verwulgen2018_dry_electodes_comfort}; (3) streaming ExG data through low-power wireless links presents privacy concerns (sensitive information could potentially be intercepted or misused) as well as limited reliability and throughput (which can compromise the integrity and usefulness of the transmitted data, especially in real-time applications) \cite{Casson2010_wearable_EEG, Tipparaju2021_BLE_in_wearables}.

This work presents \textsc{GAPses}, versatile GAP9-based smart glasses for inconspicuous, fully dry and wearable \ac{EEG} and \ac{EOG} acquisition and onboard processing. To address the above challenges, \textsc{GAPses} (1) feature a novel approach of direct electrode-electronics interfacing, resulting in an unobtrusive design that is lightweight and sleek, comparable to commodity passive glasses; (2) incorporate novel custom soft \ac{EEG} and \ac{EOG} electrodes that are designed for comfortable, fully dry acquisition of biopotentials over multiple hours of use; (3) integrate a \ac{PULP} RISC-V processor that enables on-edge processing of biopotential data, thereby eliminating the need for continuous data streaming. Thanks to PULP onboard processing, \textsc{GAPses} not only address substantial privacy concerns by reducing the risk of data interception, but also overcome issues related to bandwidth, reliability, and responsiveness of wireless connections (ensuring that the integrity and usefulness of the data are maintained, which is particularly crucial for applications requiring real-time processing).
We showcase the platform's broad applicability in measuring high-quality \ac{EEG} and \ac{EOG} data across a range of commonly used \ac{EEG} and \ac{EOG} paradigms. In particular, we validate \textsc{GAPses} in \ac{EOG}-based eye movement tasks achieving high \ac{ITR} and  in \ac{EEG}-based biometric identification, demonstrating their capabilities for end-to-end biopotential acquisition and edge processing at \ac{SoA} energy efficiency.

The main contributions are the following:
\begin{itemize}
    \item Design of a novel smart glasses platform (\textsc{GAPses}) for unobtrusive, comfortable, and secure acquisition and processing of EEG and EOG signals. The glasses weigh only \SI{40}{g} and have a compact size of \qtyproduct[product-units=single]{160 x 145}{\mm\squared}.
    \item Design of custom, fully dry soft electrodes with a direct electrode-electronics interface. Active signal buffering is provided directly at the electrodes and embedded in a sleek glasses frame.
    \item Demonstration of the application of \textsc{GAPses} on different use-cases for EOG and EEG, namely, EOG-based eye movement classification, standard EEG tasks (alpha waves, SSVEP, motor movement), and EEG-based biometrics. \textsc{GAPses} achieve an eye movement classification accuracy of \SI{96.68}{\%} on 11 classes, with an \ac{ITR} as high as \SI{161.43}{bit/min} when an accuracy of \SI{81.43}{\%} is considered sufficient. The EEG-based biometrics reach a sensitivity and specificity of \SI{98.87}{\%} and \SI{99.86}{\%}, respectively.
    \item Deployment of the proposed classification tasks on the RISC-V-based edge device, achieving an energy consumption as low as \SI{40}{\micro\joule} per inference on the EOG classification task, coupled with an average power consumption during inference of only \SI{16.02}{\milli\watt} (\SI{1.2}{\%} duty cycle). This allows for continuous operation at a total average system power of \SI{12.4}{mW} for more than \SI{22}{h} with a small \SI{75}{mAh} battery, made possible through the first-time integration of a highly versatile parallel ultra-low-power RISC-V platform in a glasses form factor for biopotentials.
\end{itemize}

Furthermore, we have open-sourced the design files of the Electrodes-interface PCB and the frame files under the permissive Solderpad v0.51 license.  The files are available online \cite{GAPses_github}.

The paper is organized as follows. Section~\ref{sec:related_work} discusses related work of wearable biopotential platforms and smart glasses, with a focus on \ac{EEG} and \ac{EOG} measurements. Section~\ref{sec:system_design} presents the design of \textsc{GAPses} with its main components: custom electrodes, frame, and acquisition and processing electronics. Section~\ref{sec:validation} validates \textsc{GAPses} with several common \ac{EEG} paradigms (alpha waves, \ac{SSVEP}, and \ac{MM}) and ocular movements (visual inspection of \ac{EOG} signals). In section~\ref{sect:EOGapp}, we showcase the application of \textsc{GAPses} for an \ac{EOG}-based task, namely, eye movement classification. In section~\ref{sect:brainmetrics_application}, we showcase the application of \textsc{GAPses} for EEG-based biometrics. Discussion and conclusion in Sect.~\ref{sect:discussion}-\ref{sec:conclusion} conclude the paper.

%% file: Sections/03_RelatedWorks.tex
\section{Related Works}\label{sec:related_work}

\subsection{Biopotential data acquisition platforms}
Examples of commercially available wireless \ac{EEG} systems include the Emotiv Insight and EPOC+ \cite{emotiv}, Neurosky MindWave \cite{neurosky}, and OpenBCI cEEGrid and Ultracortex \cite{openbci}. Of those, the Emotiv EPOC+ and OpenBCI cEEGrid are often used in research. The EMOTIV EPOC+ is designed as a headset featuring 14 \ac{EEG} channels with a \SI{14}{bit} resolution and streams the measured data at up to \SI{256}{SPS} to a PC or mobile phone. The headset can be set up quickly but does not allow for the adaptation of the channel placement and relies on a constant wireless connection to another device. OpenBCI cEEGrid stands out with its open-source approach that allows for customization in terms of the form factor (e.g., headsets or around-the-ear solutions) and electrode positions and materials. It integrates the ADS1299, a widely used \ac{AFE} for biopotential measurements that allows to measure up to 8 \ac{EEG} channels with a resolution of \SI{24}{bit} at \SI{256}{SPS} and the RFD22301 for a wireless \ac{BLE} connection.

Beyond commercial devices, several custom biopotential measurement platforms have been proposed in research. The work in \cite{Fuze_2023_EEG_headband_ADS1299} presents a headband and uses it for a depression detection task. The headband integrates three channel electrodes at the frontal lobe (Fp1, Fpz, Fp2) with bias and reference on the mastoids, and a custom PCB based on the ADS1299 for \ac{EEG} measurement is located in a box and attached to the rear side of the headband. 
Using a similar approach of integrating a dedicated \ac{AFE} for biopotential measurements, the work in \cite{Shin_2022_wearable_eeg_RHD2216} integrates the RHD2216 from Intan technologies (featuring a \SI{16}{bit} \ac{ADC}). The custom design makes use of a single \ac{EEG} channel with the electrodes placed on the forehead and on the right mastoid, and streams out the measured data via \ac{BLE} (Bot-NLE522, CHIPSEN).

A major limitation of these devices is that they lack onboard computational capabilities and need to stream raw \ac{EEG} data (possibly filtered) to a benchtop computer for processing, thereby facing the limitation of the bandwidth of the low-power data links \cite{BLE_throughput}. 
In this work, we make use of BioGAP \cite{frey_2023_BioGAP}, a \ac{SoA} biosignal acquisition and processing platform that features an ADS1298 \ac{AFE} from Texas Instruments for analog-to-digital conversion of 8 ExG channels with \SI{24}{bit} resolution and the possibility to interface both active and passive electrodes, allowing the customization of the electrode interfacing based on specific experimental needs. A key feature of BioGAP is its onboard processing capabilities powered by the GAP9 processor (GreenWaves Technologies). This enables the execution of complex \ac{DSP} tasks and \ac{ML} model inference directly on the device.
By processing data on the edge, BioGAP effectively addresses the typical bandwidth limitations and reliability concerns associated with wireless data transmission, at the same time enhancing battery lifetime.

\subsection{Smart glasses for \ac{EEG} and \ac{EOG} sensing}

A notable example of ExG glasses research prototype is e-Glass \cite{Sopic2018_e_glass_epilepsy}, equipped with four dry \ac{EEG} electrodes embedded in the glasses frame for real-time epileptic seizure detection. The hardware includes a commercially available \ac{AFE} (ADS1299, Texas Instruments) and a low-power microcontroller (STM32L151, STMicroelectronics) with limited onboard processing capabilities and wireless communication (nRF8001, Nordic Semiconductor). The authors evaluate the system's capabilities on a public dataset for seizure detection. The proposed algorithm is based on feature extraction using discrete wavelet transform (DWT) and the power in common \ac{EEG} bands with subsequent classification using a random forest classifier. Classification is based on a subset of the channels available in the dataset, corresponding to the electrode positions in e-Glass.
An embedded implementation of the proposed algorithm enables 2.71 days of operation with a \SI{570}{mAh} battery. 

The AttentivU platform  \cite{Kosmyna2019_attentivU_MIT}, on the other hand, integrates into a pair of glasses both \ac{EEG} and \ac{EOG} sensors. The device is tailored for real-time monitoring of physiological signals for attention and engagement feedback and employs dry silver electrodes: two for \ac{EEG} (TP9 and TP10), with the reference placed in the nose bridge, and two for \ac{EOG} placed in the nose pads of the glasses. The system architecture consists of preprocessing blocks for \ac{EEG} and \ac{EOG} for analog filtering and amplification, followed by a \SI{12}{MHz} microcontroller with an integrated \SI{10}{bit} \ac{ADC}). The glasses were validated by comparison with wet Ag/AgCl electrodes, and visual inspection of the signal in an \ac{AW} experiment (eyes open vs. eyes closed) for the \ac{EEG} subsystem, and execution of eye movements for the \ac{EOG} subsystem. The system can operate for \SI{5}{hours} using a \SI{150}{mAh} battery.

Another example of heterogeneous EOG-EEG sensing platform is the work in \cite{Lee2020_smart_glasses_KU_KIST_Korea},  presenting a study on 3D-printed smart glasses designed for wearable healthcare and \acp{HMI}. The smart glasses integrate electrodes based on carbon nanotube/polydimethylsiloxane composites and feature UV-responsive, color-tunable lenses for dual eyeglass and sunglass functionality. They are capable of both \ac{EEG} and \ac{EOG} sensing and integrate accelerometers for tracking of human posture and behavior. The system features an ATmega128 microprocessor (Atmel Corp.) and Bluetooth connectivity, and consumes approximately 300 mWh.

Additional examples of commercial-grade systems include both \ac{EEG}-based sensing platforms (Smith Lowdown Focus Eyewear Glasses \cite{SmithLowdownFocusEyewear}) and EOG-based sensing platforms (JINS MEME \cite{JINS_MEME}, Imec Glasses \cite{IMEC_GLASSES}). Smith Lowdown Focus Eyewear features a lightweight design with dry \ac{EEG} electrodes located at TP9 and TP10, aligning with the 10–20 \ac{EEG} system. JINS MEME integrates three-point EOG sensors on the nose pads and bridge of the glasses frame, alongside an accelerometer and a gyroscope. The Imec Glasses follow a similar electrode placement but with two additional electrodes, one on each temple of the glasses. These systems aim to monitor and enhance cognitive performance. However, they require pairing with an external device, such as a smartphone, for operation. In addition, they are limited to a single modality and cannot run custom applications directly on the glasses.

However, while the above platforms offer promising functionalities, they operate on a single biosignal modality (EEG or EOG), feature a very small number of channels, or lack sufficient onboard processing capabilities to execute more advanced ML models on the edge. 
\textsc{GAPses} overcome all these limitations, as presented in the following sections and shown in the \ac{SoA} comparison table (Table~\ref{table:soa_comparison)}, further discussed later).

%% file: Sections/04_SystemDesign.tex
\section{System Design}\label{sec:system_design}

\begin{figure*}[htb]
    \centering
    \includegraphics[width=1\textwidth, trim={0.2cm 0.1cm 0.35cm 0cm},clip]{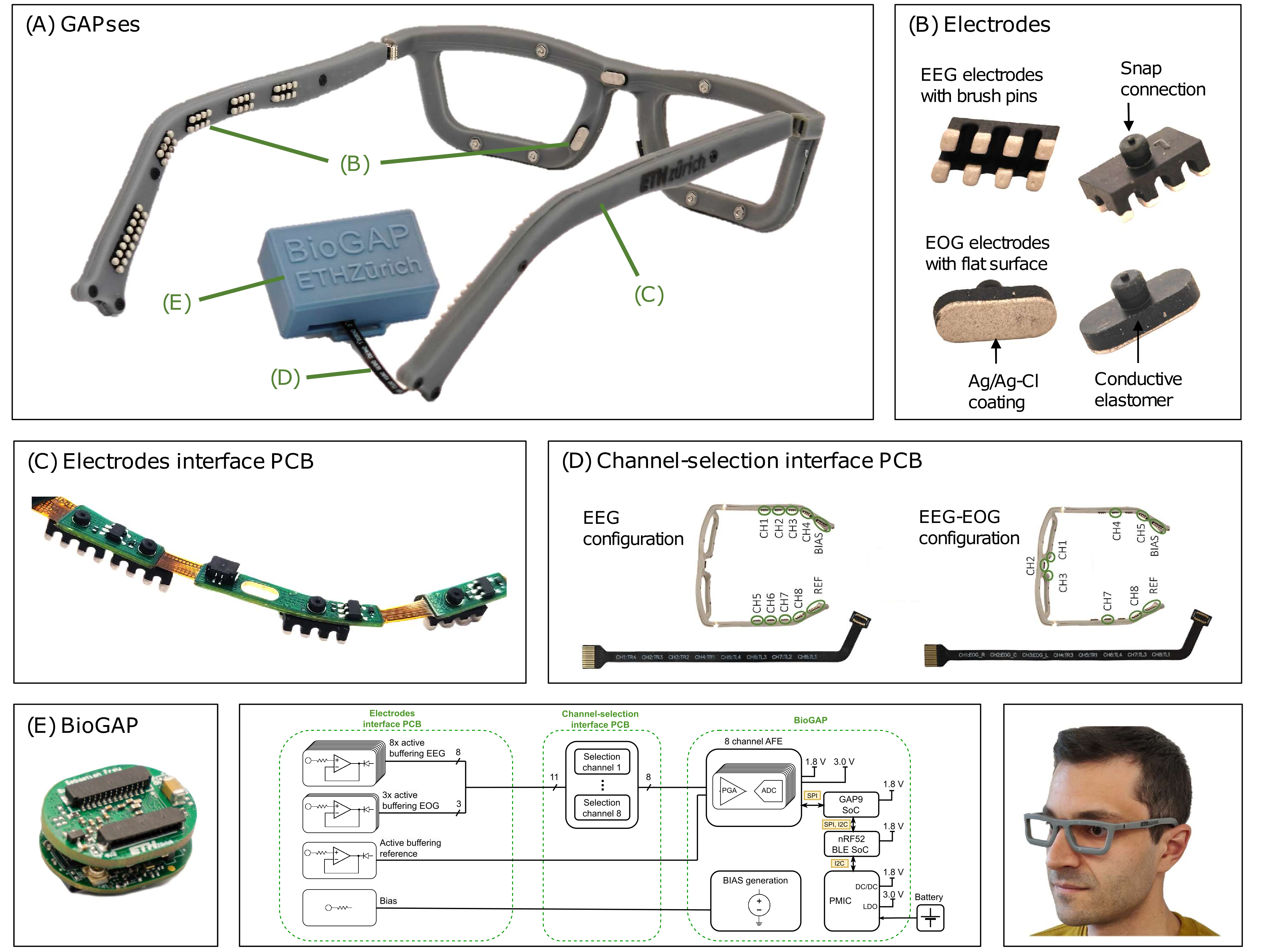}
    
    \caption{\textbf{(A)} Photo of the whole system. \textbf{(B)} Detailed view of the custom EOG (flat) and EEG (brush) electrodes. \textbf{(C)} Electrode interface PCB, embedded in the glasses' frame. \textbf{(D)} Channel-selection interface PCB, used to select which channels (among the 3 EOG and 8 EEG channels available) are interfaced to BioGAP. Two alternative versions are designed, resulting in an EEG-only configuration or hybrid EEG+EOG configuration. \textbf{(E)} BioGAP acquisition and processing platform. The bottom block diagram shows a simplified circuit diagram and the connectivity between the different PCBs and their internal key components. Bottom right: photo of a subject wearing \textsc{GAPses}}
    \label{fig:photo_system}
\end{figure*}

Figure \ref{fig:photo_system} shows the \textsc{GAPses} smart glasses, which integrate: a frame (to support the entire system mechanically), the electrodes (custom dry design by Datwyler to capture EOG and EEG signals), an electrodes interface \ac{PCB} (routing the signals from the electrodes to the acquisition device with signal buffering), a channel-selection interface \ac{PCB} (allowing to select a subset of 8 channels, enabling either a full \ac{EEG} or a combined \ac{EEG}--\ac{EOG} configuration), and BioGAP (signal acquisition and processing platform). In the following we describe each component in detail.

\subsection{Electrodes: design and placement}

The electrodes adopted in this work have been developed starting from the SoftPulse{\textregistered} technology \cite{softpulse_web} and customized to meet the requirement of the specific form factor. From a materials standpoint, the electrodes are made of two components, (i) a black electrically conductive rubber material and (ii) silver-silver chloride coating applied on the surface of the rubber part in the areas expected to be in contact with the user skin during the use of the device. The use of rubber material for the development of electrodes for bio-potential detection is a unique feature of SoftPulse and offers several advantages with respect to other dry electrodes available commercially or developed in academic environments. In fact, these electrodes can be used in dry state (i.e., without use of gel), without need for skin preparation, and are comfortable to wear for all-day use. The electrodes have been manufactured via injection molding (IM) in light of its many advantages: (i) it allows an easy upscaling due to its high automatization capabilities; (ii) IM is a very stable process, as every parameter is steered and controlled by the machine. Therefore, the reproducibility is very high and the electrodes are produced within tight tolerances and according the standard DIN ISO 3302-1. Furthermore, injection molding enables – based on the right tooling concept -  high customization degree.

All electrodes have been tested by an independent laboratory to meet the standards of biocompatibility (according to ISO 10993-5  \cite{iso10993-5}, and ISO 10993-10 \cite{iso10993-10}).

In terms of performance, SoftPulse{\textregistered} electrodes have an electrode tissue impedance (ETI) of ca. \SI{5e7}{\ohm\milli\meter\squared}, comparable to the one of metal electrodes and sintered Ag/AgCl electrodes.

For \textsc{GAPses}, two types of electrodes have been custom-designed to fit within the available space of a regular glasses frame, while still maximizing the contact area with the skin to decrease the electrode-tissue impedance. Figure~\ref{fig:photo_system} (B) shows the two electrode types, which are tailored for the acquisition of EOG and EEG signals, respectively. 

\subsubsection{EOG electrodes} 
    EOG signals are collected through three custom electrodes in the nose region. Two electrodes are placed on the nose pad, while a third one is fixed at the level of the nasion in the glass bridge, totaling 3 EOG channels (see Fig.~\ref{fig:photo_system}, A). Since the body region in contact with the EOG electrodes is not hairy, EOG electrodes have been designed with a flat surface to maximize skin contact area and minimize electrode-skin impedance. Considering the constraints coming from the glass frame, a surface area of ca. \SI{25}{\mm^2} for each electrode was considered. 
    
\subsubsection{EEG electrodes}
    EEG signals are collected using electrodes located in the back part of the glasses' arms. Specifically, \textsc{GAPses} features three \ac{EEG} channels on the temple and one channel \ac{BTE}, on each side, totaling 8 EEG channels (see Fig.~\ref{fig:photo_system}, A). Reference and bias are positioned on the temple tips at positions TP9 and TP10, respectively, according to the 10-20 reference system. Both reference and bias are obtained by connecting together two EEG electrodes, in order to maximize the contact area and stability. As the region around the ear might be hairy, EEG electrodes have been designed with prongs facing the users' skin. The presence of legs on the electrodes' surface allows contact with the skin without the need for shaving or additional skin preparation. The legs, their number, position, and orientation have been designed to allow maximum hair penetration, highest skin contact, and best comfort. In particular, the electrodes' legs have been arranged in two rows shifted by 45°. This configuration has been specifically developed thinking on how eye glasses are worn by the user, from top-front to low-back. By adopting this angle, we expect the legs to have a more efficient brush effect between the hairs, therefore achieving better hair penetration. Legs' height has also been designed specifically for integration in eyewear, with the final height of \SI{2}{\mm} being a compromise between effective hair penetration and user comfort. 
    Each EEG electrode has an area of ca. \SI{12}{\mm^2} in contact with the skin.   

\subsection{Electrodes-frames interface}

The connection between the electrodes and the electronics can be achieved by simple physical contact between the electrically conductive rubber body and the electronics. SoftPulse{\textregistered} electrodes are designed with a standard male ECG snap, and the connection is normally done by means of standard snap connectors. This configuration allows a firm and stable electrical connection, with the possibility of replacing the electrodes if needed. In this work, the use of a standard snap connector was not possible due to space limitations. In order to keep the possibility of exchanging electrodes in case of need while saving space at the same time, an ad-hoc snap mechanism has been developed. In particular, the male snap in the back of the electrodes (see Fig.~\ref{fig:photo_system}, B) has been designed together with the rigid part of the electrode-interface \ac{PCB} (Fig.~\ref{fig:photo_system}, C) where a \ac{PTH} was envisioned. 
The tight tolerances of the IM process enable perfect fit between the electrodes and the \ac{PTH} in the \ac{PCB}, which is mandatory for a stable signal quality. The \ac{PTH} metal contact allows the electrical connection.

\subsection{Glasses frame}

The design of the frame (Fig.~\ref{fig:photo_system}, A) has been developed to fulfill both aesthetic and functional requirements. 
For this reason, the design is light and thin, similar to commonly available glass frames on the market. The frame is designed in the full-rim style that completely encircles the lenses, with temple tips slightly curved behind the ears for a secure fit.

The whole frame is made of three parts, namely the front part, the right temple arm, and the left temple arm. All parts have been 3D printed with SLA technology using a semi-rigid photopolymer resin. All parts have been designed as two separated sub-parts (facing inwards and outwards, respectively, with respect to the wearer's perspective), empty inside to allow the incorporation of the electrodes interface \ac{PCB} (see below), and are assembled with screws and nuts.
The front frame has three apertures in the nose pad and bridge region to allow the insertion of the EOG electrodes. Both temple arms have six openings to fix the EEG channel electrodes and the reference / bias electrodes (three EEG electrodes on the temples, one EEG electrode \ac{BTE} electrode, two EEG electrodes connected together on the mastoids).

\subsection{Electrodes interface PCB}

Signals from the electrodes are routed to the acquisition device through a flex-rigid electrode-interface \ac{PCB} (shown in Fig.~\ref{fig:photo_system}, C). The flexible sections, integrated directly with rigid sections during manufacturing (eliminating the need for additional connectors), comprise two-layer, 0.11 mm-thick \ac{PCB}s (ENIG finish) and are employed in areas prone to bending, such as the glass legs hinges and in between the electrodes. The rigid sections, comprising four-layer, \SI{0.6}{mm}-thick \ac{PCB}s, provide a mechanical attachment and electric contact to the electrodes via \ac{PTH} with a diameter to pressure fit the snap extensions on the electrodes, allowing to attach electrodes without the need for a snap connector and hence, reducing the overall volume of the device.
Rigid sections also incorporate a buffering subsystem (one per channel), based on the AD8603 (Analog Devices), used to reduce signal interference and artifacts caused by movement while also enhancing the \ac{CMRR}. Each buffering subsystem is equipped with \SI{68}{\kilo\ohm} protection resistors to limit the current to the subject in case of a fault. Electrodes in this design rely on a single-ended configuration (monopolar montage), hence, besides individual channel leads/electrodes, the device also incorporates a reference lead/electrode and a bias lead/electrode (more details in the next section).
The electrodes interface \ac{PCB} features eight EEG channels (three temple and one \ac{BTE} channel on each side) and three EOG channels.

\subsection{BioGAP data acquisition platform and channel selection interface}\label{subsec:bioGAP_platform}

Data acquisition, processing, and wireless transmission are based on BioGAP~\cite{frey_2023_BioGAP} (Fig.~\ref{fig:photo_system}, E). 
BioGAP is a wearable and compact bio-signal acquisition and processing platform (\qtyproduct[product-units=single]{16 x 21 x 14}{\cubic\mm}) and lightweight (\SI{6}{g}). 
BioGAP encompasses two stacked \ac{PCB}s: a \emph{baseboard} and a \emph{bio-potential expansion board}. It is connected to the glasses frame via a \emph{channel-selection interface \ac{PCB}} (Fig.~\ref{fig:photo_system}, D) that selects which channel subset from the electrodes-interface PCB (out of the 3 EOG + 8 EEG channels available) is to be acquired and processed.

\subsubsection{Baseboard} The baseboard functions as the central unit for measurement control, signal processing, data handling, and power management. It hosts two \acp{SoC}: the nRF52811 from Nordic Semiconductor, which provides \ac{BLE} connectivity, and the GAP9 parallel \ac{ULP} processor by GreenWaves Technologies, tailored for \ac{DSP} computation and \ac{NN} inference with high energy efficiency~\cite{MLcommons}.
The GAP9 processor delivers up to \SI{15.6}{GOP/s} of DSP computational capability and \SI{32.2}{GMAC/s} for \ac{ML} tasks. It features automatic clock gating, voltage scaling, and adaptable and dynamic frequency scaling to optimize computational resources, minimize energy use, and facilitate extended device operation. The baseboard also incorporates \SI{256}{Mbit} of volatile memory (APS256 series, APMemory) and \SI{128}{Mbit} of non-volatile memory (MX25UW series, Macronix) for \ac{NN} weight storage, along with an \ac{IMU} (LSM6DSO, ST Microelectronics) for device interaction and motion sensing, and a \ac{PMIC} (MAX20303, Analog Devices) for efficient power regulation.

\subsubsection{Bio-potential expansion board} stacked on top of the baseboard, it extends the system's capabilities with a biopotential measurement \ac{AFE} (ADS1298, Texas Instruments). The design supports \ac{EEG} data acquisition of $8$ channels at a $24$-bit resolution. Unless otherwise stated, we operate the ADS1298 at a sampling rate of \SI{1}{kSPS} and a gain of 6 in high-resolution mode. The system is characterized by an integrated root mean square (RMS) noise of \SI{0.47}{\micro \volt} in the frequency range of \SI{0.5}{} to \SI{100}{Hz}, which aligns with the standards set by the \ac{IFCN} for recording \ac{EEG} signals in clinical settings \cite{nuwer1998ifcn}.
Additionally, the board incorporates a subsystem dedicated to evaluating electrode contact quality, ensuring the reliability and clarity of the \ac{EEG} signals collected \cite{biowolf}.

\subsection{Channel selection interface}
With BioGAP capable of processing up to 8  channels concurrently, this PCB (shown in Fig.~\ref{fig:photo_system}, D) selects which subset of channels (out of the 8 EEG + 3 EOG channels available in the frame) to use for concurrent measurement.
Two distinct channel-selection \ac{PCB}s are designed to explore two configurations:
    \begin{itemize}
        \item The \emph{EEG-only} configuration allows concurrent measurement of the six temple electrodes and two \ac{BTE} electrodes, thereby using all of the 8 EEG electrodes.
        \item The \emph{combined \ac{EEG}--\ac{EOG}} configuration selects two EEG electrodes per side as well as the three EOG electrodes on the nose bridge and pads that allow for the measurement of the vertical and horizontal EOG. The subset of EEG electrodes used in this configuration includes, on each side, the most central EEG electrode and the behind-the-ear electrode. This choice is made to keep the electrodes that are more likely to have good contact for different head sizes. 
    \end{itemize}

\subsection{System integration}

The BioGAP electronics, along with a \SI{75}{mAh} rechargeable battery, are integrated into a compact, 3D-printed enclosure, measuring \qtyproduct[product-units=single]{40 x 30 x 15}{\cubic\mm}. The box can be discretely attached to the back of the head and is connected through the flexible channel-selection interface \ac{PCB} to the electronics inside the glasses frame.
The integrated platform is assembled by inserting the \ac{EEG} electrodes into the \ac{PTH} of the electrode interface \ac{PCB}, where each electrode is paired with an opamp for active signal buffering. The \ac{EOG} electrodes are plugged into a hollow metal rod embedded in the frame, connected via wires to the active buffering stages. The electrodes interface \ac{PCB} is fully embedded within the frame and connects to the BioGAP platform through the channel-selection \ac{PCB} that allows configuration changes between the EEG only and the combined \ac{EEG}--\ac{EOG} configuration. Then, the frame is closed with lids that are secured with screws. Finally, the BioGAP box (containing BioGAP, battery, power switch, and giving access to a micro USB charging connector) can be secured on a band behind the head, which also ensures the proper fit of the glasses. The fully assembled prototype weights \SI{40}{g} and has a size of \qtyproduct[product-units=single]{160 x 145}{\mm\squared}

%% file: Sections/05_Validation.tex
\section{Validation}\label{sec:validation}

This section presents experimental results to verify the design of the prototype. These results serve as a validation step to demonstrate that \textsc{GAPses} can acquire high-quality ExG data. Here, BioGAP is operated at a sampling frequency of \SI{500}{Hz} and with a \ac{PGA} of 12.

All participants provided informed consent prior to their participation. All the experimental procedures followed the principles outlined in the Helsinki Declaration of 1975, revised in 2000.

\subsection{Validation of \ac{EOG} subsystem}
To showcase the feasibility of measuring \ac{EOG} with the given electrode configuration, a subject was instructed to wear the glasses and execute a series of saccadic eye movements: up, down, right, left, up-right, up-left, down-right, down-left, blink, and double-blink. The glasses were used with the combined \ac{EEG}--\ac{EOG} channel-selection interface PCB, as described in section~\ref{subsec:bioGAP_platform}, giving access to all three \ac{EOG} channels. Visual inspection of the \ac{EOG} signals allows to verify the validity of the measurements.

The vertical ($V_V$) and horizontal ($V_H$) \ac{EOG} signals are computed according to equations~(\ref{eq:vert_eog}--\ref{eq:horiz_eog}):

        \begin{equation}\label{eq:vert_eog}
            \begin{split}
                V_V = V_C - \dfrac{(V_R + V_L)}{2}
            \end{split}
        \end{equation}
        \begin{equation}\label{eq:horiz_eog}
            \begin{split}
                V_H = V_R - V_L
            \end{split}
        \end{equation}
        
\noindent where $V_R$, $V_L$, and $V_C$ are the signals measured from the EOG electrodes on the right side of the nose, on the left side of the nose, and the center of the glass bridge, respectively.

\begin{figure}[t]
\centerline{\includegraphics[width=1.0\columnwidth]{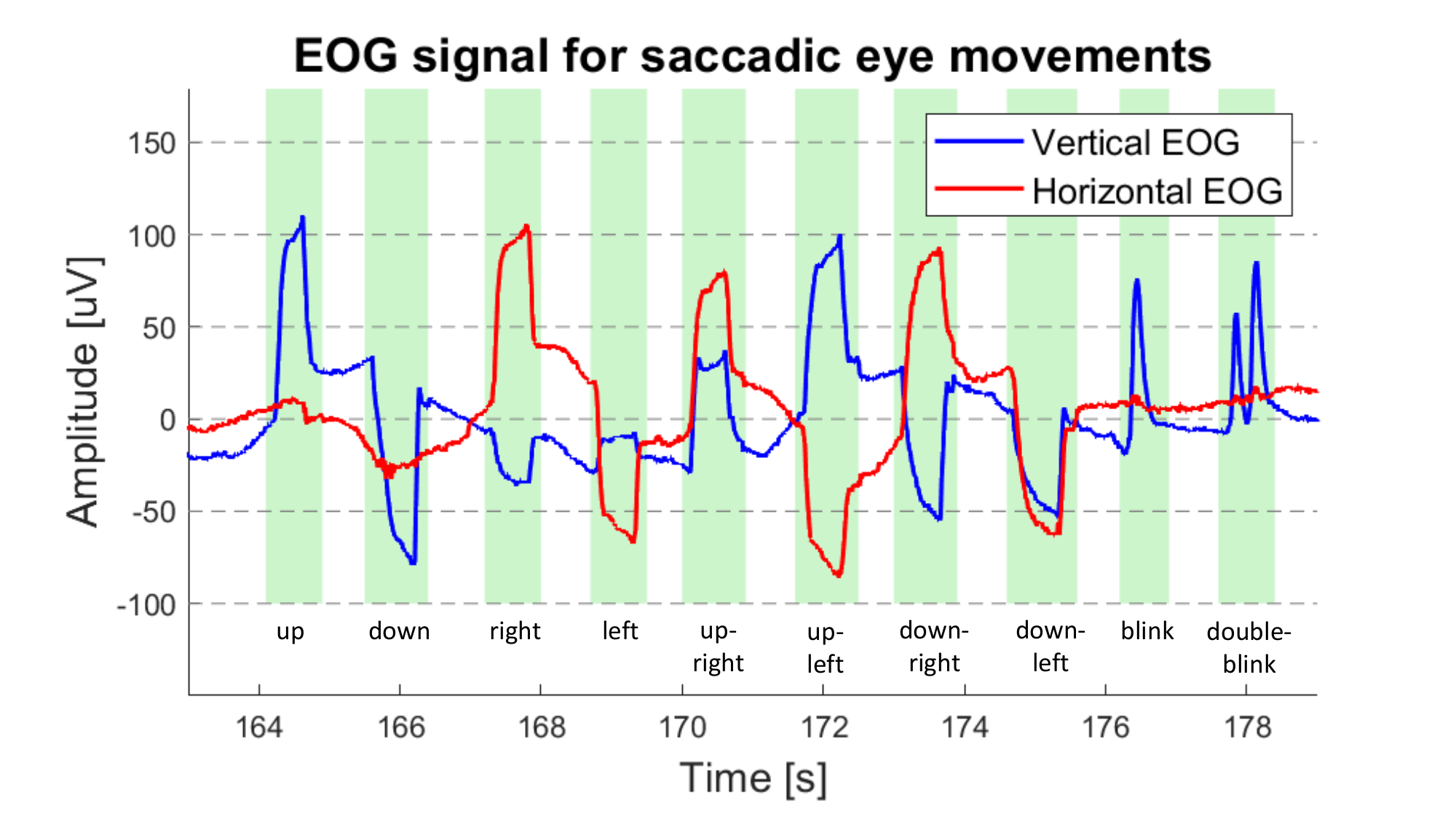}}
\caption{Measurement of the horizontal and vertical EOG signal while different eye movements are performed.}
\label{fig:eog_time}
\end{figure}

The offset and signal drift are removed by subtracting the running mean with a window size of \SI{2}{s} and the signal is subsequently smoothed with a 10-th order IIR Butterworth low-pass filter with a \SI{40}{Hz} cutoff frequency. Figure~\ref{fig:eog_time} shows a series of eye movements, where the subject is instructed to look from the center to the respective direction (e.g., to the right, left, etc.) and then right back to the center. Purely vertical eye movements (up, down) show a strong response of the vertical \ac{EOG} in the respective direction, while the horizontal \ac{EOG} amplitude stays low. Conversely, purely horizontal eye movements (right, left) show a strong response in the horizontal \ac{EOG} component, while the vertical signal remains low. Diagonal eye movements (up-right, up-left, down-right, and down-left) result in the respective combination of horizontal and vertical \ac{EOG} signal. Finally, blink and double-blink result in one and two short peaks in the vertical \ac{EOG} signal. The distinct signal shape of each eye movement confirms the feasibility of acquiring high-quality \ac{EOG} signals with the given electrode positions and distinguishing between the chosen eye movements, while the fast sequence of distinct movements suggests the feasibility of implementing a \ac{HMI} with a high \ac{ITR} (Sect.~\ref{sect:EOGapp} provides more insights into EOG-based applications).

\subsection{Validation of \ac{EEG} subsystem}
\subsubsection{Alpha waves}

\begin{figure}[b]
\centerline{\includegraphics[width=1.0\columnwidth]{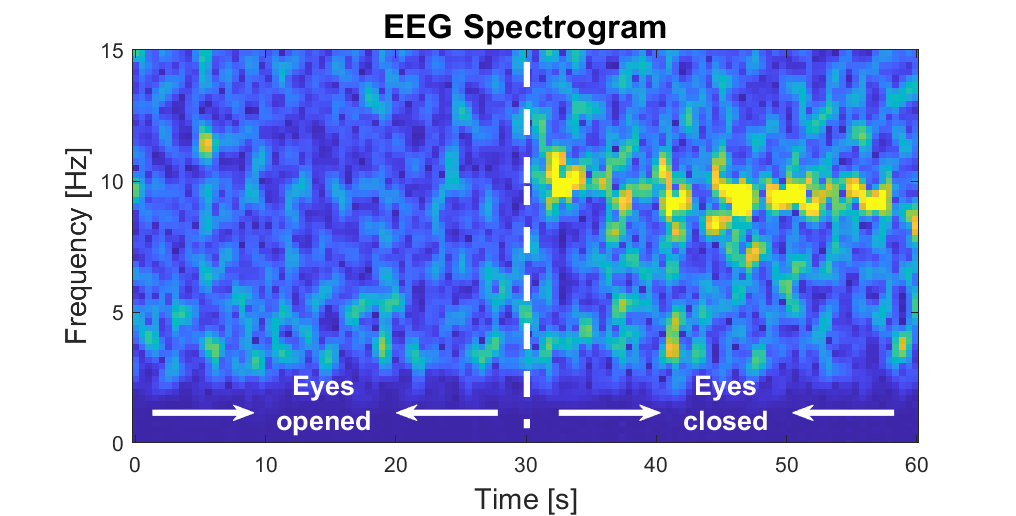}}
\caption{Spectrogram of the EEG system evaluation by performing alpha waves measurement in the eyes open vs eyes closed experiment.}
\label{fig:alpha_waves}
\end{figure}

A subject was tested in an alpha waves experiment using the glasses in the \ac{EEG}-only configuration.
The subject was instructed to alternate between closed and open eyes states for 30 seconds each, as this paradigm is commonly used as a qualitative marker for EEG-based systems in both medical and \ac{HMI} applications \cite{turk2015changes}.
Figure~\ref{fig:alpha_waves} shows the signal spectrogram (1024 samples=\SI{2.048}{s} windows with a 768 samples=\SI{1.536}{s} overlap) of the channel 4 in the \ac{EEG} configuration (CFR. Fig.~\ref{fig:photo_system}, D). As expected, a strong increase in the energy component for the alpha band is evident when the subject was instructed to close the eyes.

\subsubsection{\ac{SSVEP}}
\begin{figure}[t]
\centerline{\includegraphics[width=1.0\columnwidth]{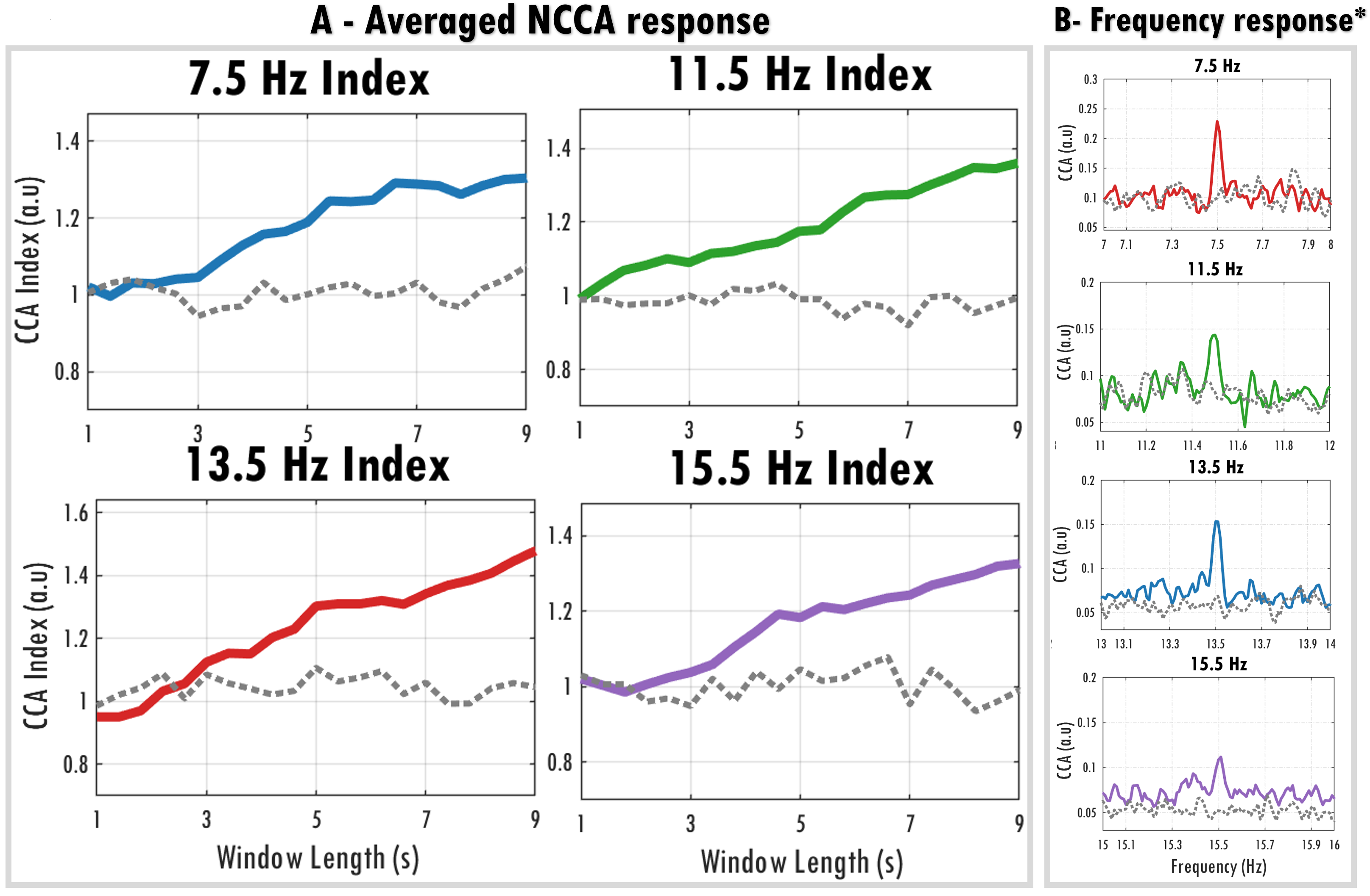}}
\vspace{0cm}
\caption{(A) NCCA of the \ac{SSVEP} experiment showing the response at different window lengths (average values across five subjects). (B) corresponding CCA for one subject.}
\vspace{0cm}
\label{fig:ssvep_cca}
\end{figure}

\ac{SSVEP} is a frequency-and-phase-locked EEG response to repetitive visual stimuli \cite{zhu2010survey}, frequently employed as a control paradigm in \ac{SoA} \ac{BCI}s. We evaluated the glass's performance for five subjects. Subjects were sitting in front of a 14-inch computer screen at approximately \SI{60}{cm} distance. Stimuli, consisting of sinusoidal on-off patterns with 100\% contrast, were presented sequentially on the screen. Pattern's frequency include \SI{7.5}{Hz}, \SI{11.5}{Hz}, \SI{13.5}{Hz}, and \SI{15.5}{Hz}, and were presented in a random order, across three repetitions. Each frequency was presented for 25 seconds, followed by a 10-second rest period to mitigate visual fatigue. 
EEG response is measured through the \ac{NCCA} \cite{kartsch2022efficient}. \ac{NCCA} extends the \ac{CCA} by focusing on the detection of a specific "peak" frequency within EEG data.  \ac{NCCA} hence, provides the ratio of the CCA response at a target frequency to the average response at two adjacent frequencies. In this work, \ac{NCCA}\footnote{Side frequencies for NCCA calculation are always ±0.2 Hz from the target frequency.} is computed on data segments from each trial, with results averaged among all available trials. The results for each target frequency (colored continuous lines) are presented in Fig~\ref{fig:ssvep_cca} (A), with increasingly larger \ac{CCA} evaluation windows. \ac{NCCA} is also computed for rest trials (grey dashed lines) to provide evidence of the correct operation of the \ac{NCCA} index algorithm. For all frequencies, a window of \SI{3}{s} is sufficient to identify SSVEP response with sufficient confidence\footnote{Empirical data show that \ac{NCCA} values above 1.1 significantly correlate to the presence of EEG SSVEP response.}. For reference, Fig~\ref{fig:ssvep_cca} (B) also presents the frequency response (based on the \ac{CCA} algorithm) for all four frequencies for subject 1, each denoting clear power peaks that are always above power values of neighbor frequencies and the rest segments.

\subsubsection{\ac{MM}}
The glasses were further validated on a \ac{MM} protocol, in light of its popularity among BCI paradigms and to verify signals originated from the motor cortex. Five subjects performed right and left-hand movements (finger tapping~\cite{chang2018_bcitutorial}) interleaved by rest periods while EEG data were recorded. Trial duration (gesture+rest) is, on average, 10s, with gestures lasting 4s and rests of random duration between 5 and 6 seconds. Subjects followed visual instructions on a screen placed ca. 1\,m away, and each instruction synchronized with the \ac{EEG} data through a digital triggering system on BioGAP. \ac{MM}classification performance is assessed through MI-BMInet~\cite{wang2023enhancing}, a lightweight, embedded \ac{CNN} that demonstrated \ac{SoA} classification accuracy for this task. 
The accuracy obtained through the classifications is reported on the basis of a rolling window \ac{CV} on a validation set. We train the models (subject-specific) for 500 epochs using cross-entropy loss, Adam optimizer (lr=0.001), and a batch size of 64.

Before training and inference, \ac{EEG} data are preprocessed with a notch filter at 50\,Hz and a $4^{th}$ order IIR band-pass filter at 0.5--100\, Hz and downsampled by a factor of 2, obtaining 950 samples in time. Table \ref{tab:MM_val} presents the 2-class (left vs. right) and 3-class (left vs. right vs. rest) classification results of MI-BMInet for all five subjects and the resulting average. Peak classification performance is achieved by S1, with 67\% and 49\% of accuracy for the 2-class and 3-class classification, respectively. The reported averaged values are also above chance (11.2\% and 8.92\% for 2-class and 3-class, respectively), indicating the device's ability to retrieve basic \ac{MM} events.

\begin{table}[]
    \centering
    \caption{Accuracy scores for the \ac{MM} classification task.}
    \begin{tabular}{lcc}
    \toprule
    \textbf{Subject} & \multicolumn{1}{c}{\textbf{2 class}} & \multicolumn{1}{c}{\textbf{3 class}} \\
             & \multicolumn{1}{c}{\textbf{accuracy [\%]}} & \multicolumn{1}{c}{\textbf{accuracy [\%]}} \\
    \midrule
    S0 & 57 & 37.33 \\
    S1 & 67 & 49.33  \\
    S2 & 59 & 41.33 \\
    S3 & 62 & 40 \\
    S4 & 61 & 43.33 \\
    \bottomrule
    \textbf{Average} & \textbf{61.2} & \textbf{42.26} \\
    \end{tabular}
    \label{tab:MM_val}
\end{table}

%% file: Sections/06_Applications.tex
\section{EOG application: automated saccadic eye movement classification}
\label{sect:EOGapp}

Sect.~\ref{sec:validation} validated the acquisition of EOG signals with \textsc{GAPses} based on visual inspection. In this section, we automate the eye movement classification task by deploying on the device an energy-efficient classification pipeline, demonstrating a competitive \ac{ITR} performance at ultra-low power. These results demonstrate that \textsc{GAPses} are well-suited for wearable and fast Human-Machine interactions, e.g. for assisted spelling or gaming.

\subsection{Data acquisition protocol}\label{subsec:EOG_data_acquisition_protocol}
Five subjects participated in an eye movement experiment where the horizontal and vertical \ac{EOG} was measured while the subjects were seated in a stationary position. A Python script (based on Psychopy) provided the subject with visual instructions on a monitor on which eye movement should be performed. The measured raw data were streamed to a Java GUI and saved alongside the ground truth labels provided by the Psychopy script. The subjects performed the eye movements freely, with no specific fixation point to guide them. While this might lead to a slight decrease in accuracy, it is expected to result in a more diverse dataset that improves the robustness of the model in realistic scenarios and less constrained settings. The dataset encompasses 11 classes (up, down, right, left, up-right, up-left, down-right, down-left, blink, double-blink, rest), with two sessions per subject and removal/repositioning of the glasses between sessions (to introduce inter-session variability mimicking realistic use-case scenarios, where end users do not wear the device all the time). Each session involved recording 25 trials, each with a duration of \SI{2}{s},  for every class of eye movement, resulting in 50 samples per class for each subject.

\subsection{Data processing pipeline and classification model}\label{subsec:EOG_processing_pipeline}
Firstly, we calculate the vertical and horizontal EOG signals using equations~\ref{eq:vert_eog}~and~\ref{eq:horiz_eog}, respectively. Then, we apply a bandpass filter between 0.5 and 40Hz and apply a moving average filter on the signal. The moving average filter uses a window size of 2 seconds and only takes into account past data for averaging removal. We segment the measured data stream into samples corresponding to individual eye movements using the time-synchronized ground-truth labels.
The classification task is based on a modified version of the \textsc{EpiDeNet} network~\cite{thorir_biocas_2023}, adapted to classify EOG signals. The modified network, detailed in Table~\ref{tab:EpiDeNet_Table}, includes a parameterization of the kernel size for the last MaxPool layer to be either 1 for EOG signals or 4 for EEG signals.
\input{tables/architecture}

\subsection{Results}

\begin{figure}[b]
\centerline{\includegraphics[width=1.0\columnwidth]{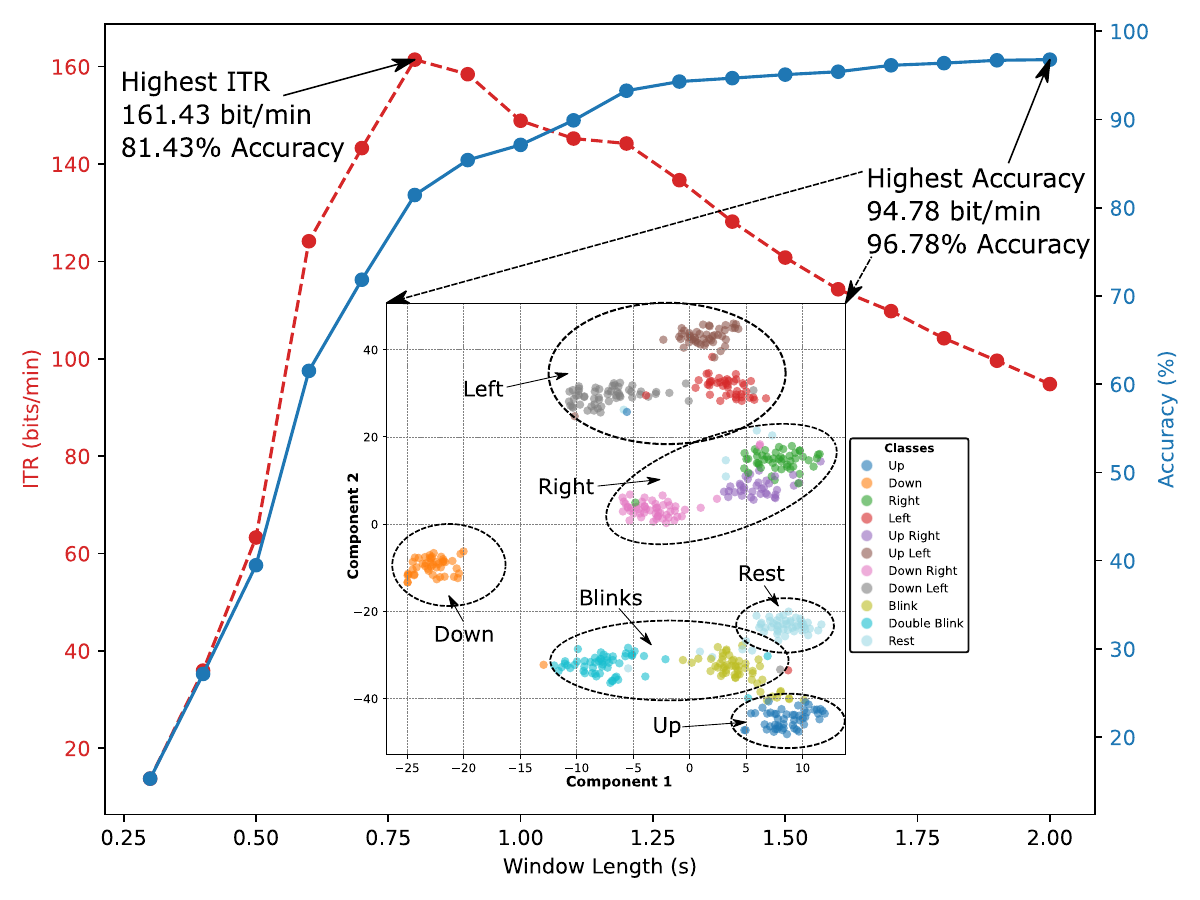}}
\vspace{-0.3cm}
\caption{Averaged Accuracy and ITR values for the subject-specific models when considering different fractions of the 2-second windows. Inset: t-SNE visualization of the highest accuracy point.}
\vspace{-0.4cm}
\label{fig:eog_itr_acc_tsne}
\end{figure}

We evaluated our approach via a 5-fold cross-validation. We examined both a 'global' model, which was trained on all subjects, and a subject-specific model. The average accuracy of the global model was $94.91\%$ (not shown), while the subject-specific model achieved an average accuracy of $96.78\% \pm{2.44}\%$. Fig.~\ref{fig:eog_itr_acc_tsne} (blue curve) shows how the subject-specific accuracy changes when only a fraction of the 2-second windows is considered. The maximum accuracy is achieved when the entire 2-second window is used.

Additionally, we visualized the features the \textsc{EpiDeNet} model outputs from the last convolutional layer using t-SNE~\cite{van2008visualizing}. The inset in Fig.~\ref{fig:eog_itr_acc_tsne} shows the 2-dimensional representation of the feature space explored by \textsc{EpiDeNet}. Similar movements (such as double blinks and single blinks) are grouped together, and most errors in the subsequent classification derive from the selected boundary between these classes.

The \ac{ITR} is computed based on equation~\ref{eq:ITR}:
\begin{equation}\label{eq:ITR}
ITR = \frac{60(\log_2 M + P\log_2 P + (1-P)\log_2 \frac{1-P}{M-1})}{T}
\end{equation}

with $T$ being the window length, $P$ being the accuracy of the classification, and $M$ being the number of different tasks (11 in our case).
Fig.~\ref{fig:eog_itr_acc_tsne} (red line) shows how the \ac{ITR} changes with the considered window length. The maximum \ac{ITR} is achieved for $T = \SI{0.8}{s}$, with a corresponding performance (accuracy) of $P = 0.814$, and is equal to ITR$_{\textrm{max}}=$ \SI{161}{bit/min}. 

\subsection{EOG Artifact robustness}

We evaluated the impact of motion artifacts as a further step towards adapting GAPses for real-life conditions. To isolate the effect of motion artifacts from other variables, we designed an additional experiment with data collected from a subject under two different conditions: rest and walking. 
To enable performing experiments during walking, the Psychopy script delivered audio (instead of visual) instructions to the test subject on which eye movements to perform. The number of classes and the number of trials performed per class are the same as in the protocol described above. 

\begin{figure}[bth]
\centerline{\includegraphics[width=0.8\columnwidth]{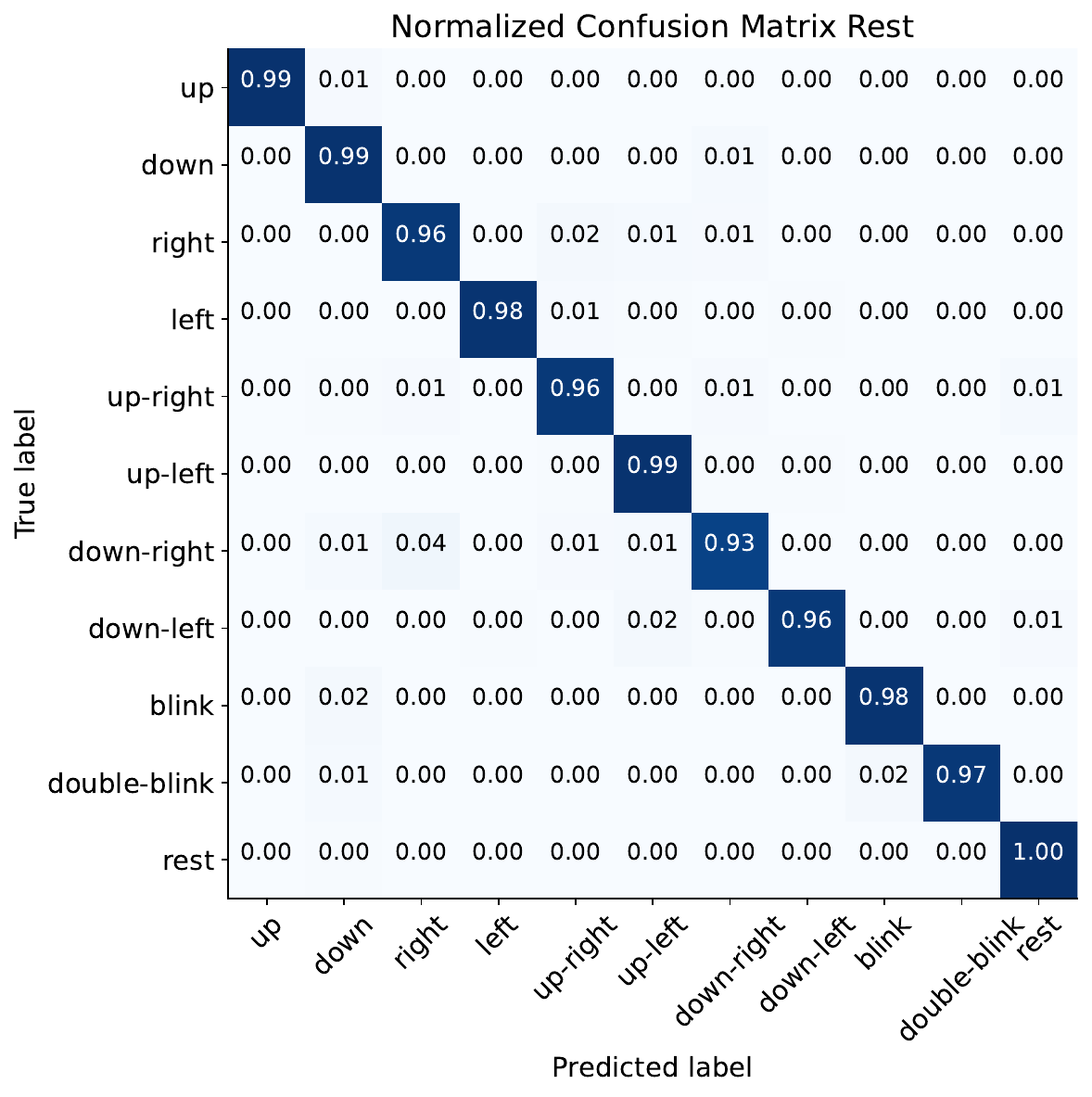}}
\vspace{0.2cm}
\centerline{\includegraphics[width=0.8\columnwidth]{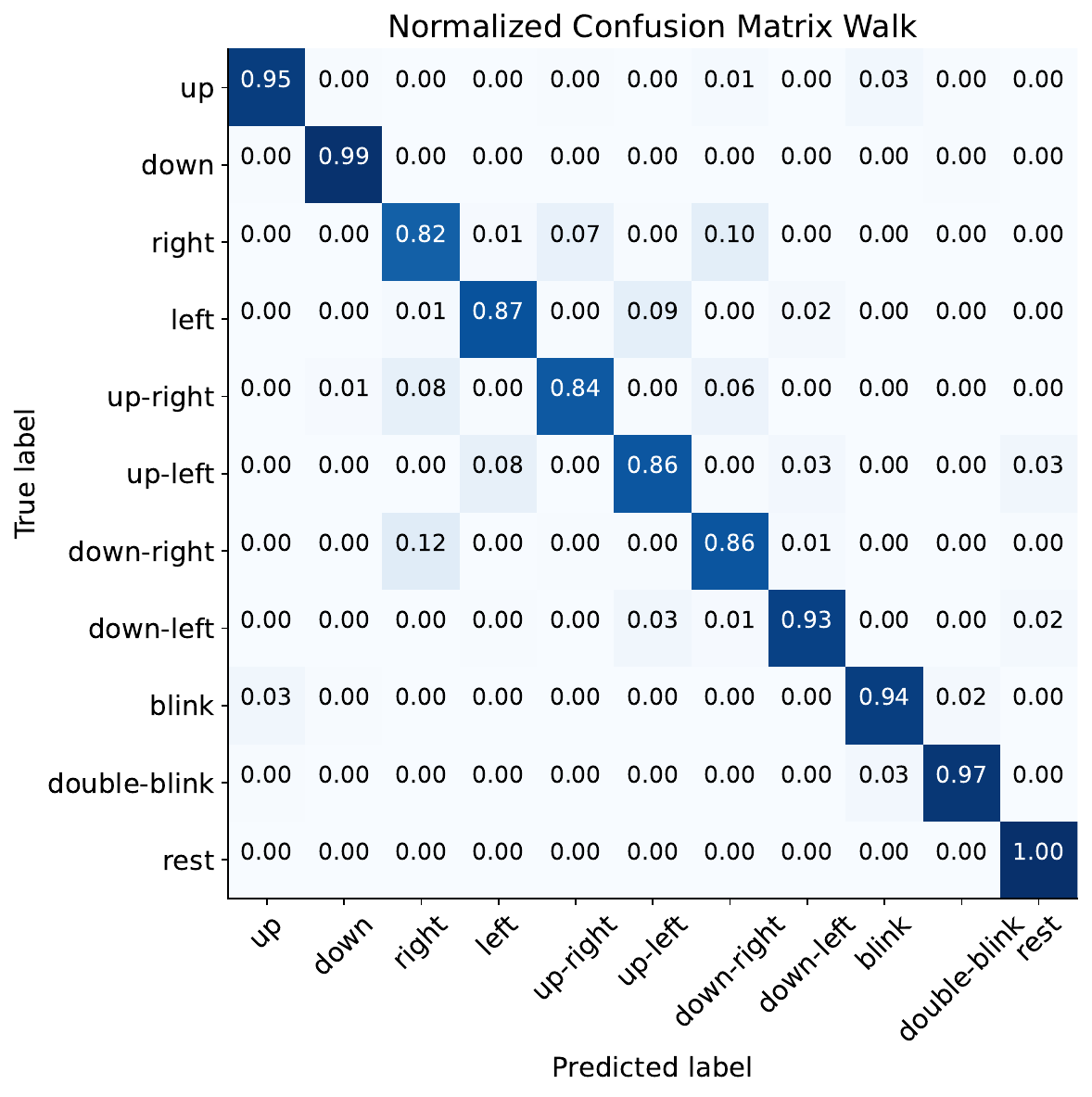}}
\vspace{-0.3cm}
\caption{Confusion matrices of the EOG saccadic eye movement experiment of rest (top) and walking condition (bottom).}
\vspace{-0.4cm}
\label{fig:eog_artifact_confusion_matrix}
\end{figure}

The classification results for the two test conditions (rest and walking) are shown in Fig~\ref{fig:eog_artifact_confusion_matrix}.
Side and diagonal movements (right/up-right/down-right, left/up-left/down-left) appear as the most affected by motion artifacts during the walking condition. This result is in agreement with intuition, as during walking the user might have a less precise control on the specific diagonal angle movements, considering the need to keep focus on the road ahead while moving. Despite this, the decrease in classification accuracy while walking (\SI{91.2}{\%}) compared to the rest condition (\SI{97.4}{\%}) remains small, proving that \textsc{GAPses} can be used also while moving, under realistic conditions.

\subsection{Embedded implementation}
\label{sect:deploymentEOG}
We deploy our model on the GAP9 platform. We employ Quantlab~\cite{quantlab} to convert neural networks into the INT-8 format (a step needed for adapting the models to low-power embedded platforms). Additionally, the deployment process is facilitated by DORY~\cite{burrello2021dory}, a specialized tool that autonomously generates C code tailored for managing the two-tier memory architecture—L1 and L2 memory—found in PULP-based systems. DORY optimizes memory utilization, ensuring that our deployment maximizes efficiency and effectiveness in resource-limited environments.
Table \ref{tab:results:summary} (EOG columns) shows the result of the embedded implementation when considering two alternative input sizes (either 500 or 1000 samples, for the two EOG signals $V_H$ and $V_V$), achieving a peak energy efficiency of 12.24 GMAC/s/W, an energy per inference as low as 0.040 mJ, and an average power consumption during inference of only 16.02 mW.

\subsection{Comparison to SoA for EOG-based eye movement classification}

\textsc{GAPses} achieve competitive performance in eye movement classification (peak of 96.78\%), also when compared to conventional acquisition setups. 
In fact, \cite{ShangLinWu2013} demonstrated an accuracy of 88.59\% for eight distinct eye movements and single blinks, and \cite{OBard2018} achieved an accuracy of 96.9\% for six saccades classes coupled with a 97.33\% eye-blink detection accuracy.
Considering alternative wearable devices, \cite{Barbara2016} used the JINS MEME smart glasses to develop an \ac{EOG} speller and classified 16 distinct eye movements and a rest class using a thresholding algorithm, resulting in a classification accuracy that is significantly lower (73.78\%) than the accuracy achieved in this work. 

The achieved \ac{ITR} also proved to be competitive. Compared to the recent work of \cite{he2017single}, which achieved an \ac{ITR} of 38 bit/min with a single-channel EOG, \textsc{GAPses} offers an approx. $4\times$ higher performance. Additionally, our \ac{ITR} also surpasses the 108.63 bits/min achieved by \cite{zhang2023online} with a hybrid EOG-SSVEP approach. Furthermore, our population size is comparable or only slightly smaller compared to the SoA work (10 subjects \cite{Barbara2016}, 8 subjects \cite{he2017single}, 10 subjects \cite{zhang2023online}.

These results demonstrate that \textsc{GAPses} coupled with \textsc{EpiDeNet} is a leading-edge solution for low-power and fast EOG-based HMIs.

\section{EEG Application: BrainMetrics}\label{sect:brainmetrics_application}

Sect.~\ref{sec:validation} validated the acquisition of EEG signals with \textsc{GAPses} based on the most common paradigms (alpha waves, SSVEP, MM). In this section, we demonstrate the application of \textsc{GAPses} for EEG-based biometrics (BrainMetrics), i.e., to identify the glasses' owner through their unique brainwave patterns. This feature allows for a more personalized and secure interaction with the device. By verifying the user's identity with EEG-based biometrics, BrainMetrics ensures that only the authorized user can access the device’s advanced features, such as the EOG-based interaction (previous section).

\subsection{Data acquisition protocol}

Six healthy subjects participated in three separate sessions spread over multiple days, ensuring a diverse dataset with multiple repositioning of the glasses. During each session, we conducted half an hour of \ac{EEG} data acquisition with the \ac{EEG}-only configuration of the glasses (see Fig.~\ref{fig:photo_system}, D) that allows concurrent measurement of eight \ac{EEG} channels along the temples and \ac{BTE}. To capture a diverse set of data reflective of typical daily activities, subjects could engage in their normal routines, such as working on a computer, reading, and watching television. However, to mitigate the impact of motion artifacts on the data quality, we requested that all participants remain seated throughout the experiment. Additionally, subjects were advised to minimize talking to reduce further potential disruptions in the \ac{EEG} recordings.

\subsection{Data processing pipeline and model architecture}
Data are first band-pass filtered within the 0.5 Hz to 100 Hz range, and subsequently, we apply a 50 Hz notch filter to remove any electrical noise. A moving average filter with a duration of 0.5 seconds is utilized to enhance the uniformity of the data. The segmented data are organized into 4-second intervals, each assigned a label from 0 to 5, indicating the specific subject the data originated from.
Regarding the model architecture, we employ \textsc{EpiDeNet}~\cite{thorir_biocas_2023} in light of its robustness and energy-efficient performance, particularly on the GAP9 platform.

\subsection{Results}
We explore a subject-specific model approach, i.e., we train a dedicated model for each subject, specialized in identifying whether the EEG signature belongs to the particular subject it's trained on. Hence, the model operates in a 2-class modality, classifying ''owner" vs ''non-owner" of the glasses. We segment the data and employ a 5-fold cross-validation technique to obtain a reliable assessment of the model's precision. We evaluate the outcomes for the two available EEG-channel configurations: employing all eight electrodes (EEG configuration) versus using just four electrodes (combined EEG - EOG configuration) for the classification task. This comparison demonstrates the methodology's robustness in maintaining high accuracy levels, even when only a partial set of channels is utilized.

Table~\ref{tab:brainmetrics:individual} shows the classification results and reveals a high average specificity (99.86), coupled with a high sensitivity (98.87). These findings indicate that a subject-specific model constitutes an effective strategy for BrainMetrics applications.

\begin{table}[]
    \centering
    \caption{\textsc{BrainMetrics} Accuracy, Sensitivity and, Specificity scores for subject-specific models.}
    \begin{tabular}{lrrrr}
    \toprule
    Metric & Subject & Mean (8 channels) & Mean (4 Channels) \\
    \midrule
    Accuracy & S0 & 99.91 $\pm$ 0.04 & 99.93 $\pm$ 0.05 \\
    Accuracy & S1 & 99.57 $\pm$ 0.15 & 99.57 $\pm$ 0.15 \\
    Accuracy & S2 & 99.89 $\pm$ 0.04 & 99.89 $\pm$ 0.04  \\
    Accuracy & S3 & 99.84 $\pm$ 0.14 & 99.84 $\pm$ 0.14 \\
    Accuracy & S4 & 99.32 $\pm$ 0.16 & 99.30 $\pm$ 0.12  \\
    Accuracy & S5 & 99.92 $\pm$ 0.07 & 99.91 $\pm$ 0.10  \\
    Accuracy & mean & 99.74 $\pm$ 0.11 & 99.74 $\pm$ 0.11  \\
    \midrule
    Sensitivity & S0 & 99.74 $\pm$ 0.19 & 99.79 $\pm$ 0.22 \\
    Sensitivity & S1 & 99.80 $\pm$ 0.20 & 99.80 $\pm$ 0.21 \\
    Sensitivity & S2 & 99.36 $\pm$ 0.25 & 99.27 $\pm$ 0.24  \\
    Sensitivity & S3 & 98.07 $\pm$ 1.83 & 98.07 $\pm$ 1.83 \\
    Sensitivity & S4 & 96.76 $\pm$ 0.96 & 96.51 $\pm$ 0.96  \\
    Sensitivity & S5 & 99.49 $\pm$ 0.29 & 99.49 $\pm$ 0.29 \\
    Sensitivity & mean & 98.87 $\pm$ 0.87 & 98.82 $\pm$ 0.87  \\
    \midrule
    Specificity & S0 & 99.96 $\pm$ 0.05 & 99.98 $\pm$ 0.04 \\
    Specificity & S1 & 99.48 $\pm$ 0.14 & 99.48 $\pm$ 0.14 \\
    Specificity & S2 & 99.98 $\pm$ 0.04 & 100.00 $\pm$ 0.00 \\
    Specificity & S3 & 99.97 $\pm$ 0.04 & 99.97 $\pm$ 0.04 \\
    Specificity & S4 & 99.79 $\pm$ 0.09 & 99.81 $\pm$ 0.11 \\
    Specificity & S5 & 99.97 $\pm$ 0.07 & 99.95 $\pm$ 0.10  \\
    Specificity & mean & 99.86 $\pm$ 0.08 & 99.87 $\pm$ 0.09  \\
    \bottomrule
    \end{tabular}
    
    \label{tab:brainmetrics:individual}
\end{table}

\subsection{Embedded implementation}
We use the same deployment procedure presented in Sect.~\ref{sect:deploymentEOG} for the EOG embedded implementation.
Table~\ref{tab:results:summary} (EEG column) shows the achieved performance for the deployment of \textsc{BrainMetrics} on \textsc{GAPses}. We demonstrate an energy efficiency as high as 31.64 GMAC/s/W, coupled with an energy per inference of only 0.121 mJ and average power consumption during inference as low as 26.54 mW.

Table~\ref{tab:power_cunsumption_breakdown} provides a detailed breakdown of the power consumption for both end-to-end applications, covering individual circuit parts and the complete system. Most of the power consumption is attributed to the analog signal conditioning and sampling (\ac{AFE} and active electrode circuitry), as well as data handling (data reception of GAP9 and BLE connection, and control code of the nRF52811). For the EOG application, the contribution of the \ac{AFE} and active electrodes is smaller since only three channels are used. Furthermore, the power needed for running inference on the edge (\SI{0.12}{mW}, \SI{0.20}{mW}, and \SI{0.61}{mW}, for the EOG application with input sizes of 2x500 and 2x1000, and the EEG application, respectively) is minimal compared to other contributors.

\begin{table}[ht!]
\renewcommand{\arraystretch}{0.95}
  \centering
  \caption{Implementation of the EOG and BrainMetrics network on GAP9}\label{tab:results:summary}
  \vspace{-.2cm}
  {
    \footnotesize
    \begin{tabular}{lrrr}
      \toprule
      Network & \multicolumn{1}{c}{\textsc{EpiDeNet$^*$}} & \multicolumn{1}{c}{\textsc{EpiDeNet$^*$}} & \multicolumn{1}{c}{\textsc{EpiDeNet$^*$}} \\
      Data & \multicolumn{2}{c}{EOG} & \multicolumn{1}{c}{EEG} \\
      \cmidrule(r){1-1} \cmidrule(r){2-3} \cmidrule(r){4-4}
      Platform (MCU) & \multicolumn{3}{c}{GAP9 (1+9$\times$RISCY @240\,MHz)}\\
     \cmidrule(r){2-4}
      Deployment framework & \multicolumn{3}{c}{Quantlab/DORY} \\ 
     \cmidrule(r){2-4}
     Input size & $2\times500$ & $2\times1000$ & $8\times2000$\\
     \cmidrule(r){2-2}\cmidrule(r){3-3}\cmidrule(r){4-4}
      MACs & 259,856 & 484,752 & 3,844,000 \\
      \cmidrule(r){1-1} \cmidrule(r){2-2}\cmidrule(r){3-3} \cmidrule(r){4-4}
      Time/inference [ms]   & \textbf{1.50} & \textbf{2.47} & \textbf{4.58} \\
      Throughput [MMAC/s] & \textbf{173.47} & \textbf{196.10} & \textbf{839.67} \\
      MACs/cycle & \textbf{0.72} & \textbf{0.82}& \textbf{3.50} \\
      Power during inf. [mW] & \textbf{16.28} & \textbf{16.02} &\textbf{26.54} \\
      Energy/inference [mJ]   & \textbf{0.024} & \textbf{0.040} & \textbf{0.121} \\
      En. eff. [GMAC/s/W] & \textbf{10.66} & \textbf{12.24} & \textbf{31.64} \\
      \bottomrule
    \end{tabular}
  }
  \label{tab:network_deployed_SoA}

  \vspace{-.3cm}
\end{table}

\begin{table}[ht!]
\renewcommand{\arraystretch}{0.95}
  \centering
  \caption{Power breakdown of the EOG and BrainMetrics end-to-end applications}\label{tab:results:power_breakdown}
  \vspace{-.2cm}
  {
    \scriptsize
    \begin{tabular}{lrrr}
      \toprule
      Application & \multicolumn{2}{c}{Eye movement class.} & \multicolumn{1}{c}{Brainmetrics} \\
      \cmidrule(r){1-1} \cmidrule(r){2-3} \cmidrule(r){4-4}
     Input size & $2\times500$ & $2\times1000$ & $8\times2000$\\
     \cmidrule(r){2-2}\cmidrule(r){3-3}\cmidrule(r){4-4}
      \cmidrule(r){1-1} \cmidrule(r){2-2}\cmidrule(r){3-3} \cmidrule(r){4-4}
      Computation [mW] $^{(1)}$   & 0.12 (\SI{0.97}{\%}) & 0.20 (\SI{1.61}{\%}) & 0.61 (\SI{3.13}{\%}) \\
      Data handl. \& BLE [mW] & 3.5 (\SI{28.4}{\%}) & 3.5 (\SI{28.2}{\%}) & 3.5 (\SI{17.9}{\%}) \\
      AFE \& act. electr. [mW] & 6.2 (\SI{50.3}{\%}) & 6.2 (\SI{50.0}{\%})& 11.5 (\SI{58.9}{\%}) \\
      Regulator ineff. [mW] & 2.5 (\SI{20.3}{\%}) & 2.5 (\SI{20.2}{\%}) & 3.9 (\SI{20.0}{\%}) \\
      Total power [mW] & \textbf{12.3 (\SI{100}{\%})} & \textbf{12.4 (\SI{100}{\%})} & \textbf{19.5 (\SI{100}{\%})} \\
      \bottomrule
    \end{tabular}
    \begin{tablenotes}\scriptsize
        \item[] $(1)$ Average GAP9 computation power when using a window overlap of 200~ms (5 inferences per second)
    \end{tablenotes}
  }
  \label{tab:power_cunsumption_breakdown}

  \vspace{-.3cm}
\end{table}

\input{tables/SoA_comparison}

\subsection{Comparison to SoA for EEG-based biometrics}

Our analyses are based on six subjects as a feasibility study of BrainMetrics with \textsc{GAPses}. The number of subjects is consistent with multiple related works that also make use of small population size and a comparable number of EEG channels \cite{touyama2009eeg,Yu2019,Ashby2011}.
Since most of the related works report accuracy metrics, we base the following comparison on the mean accuracy achieved by 
\textsc{GAPses}, i.e., 99.74\% (for eight channels).

Our accuracy is comparable to that achieved with EEG-cap setups by \cite{bak2023user} (98.97\% accuracy, based on a motor-imagery paradigm) and \cite{vadher2024eeg} (99.86\% validation accuracy, based on a multi-task approach).
Additionally, our classification accuracy also surpasses the performance of works employing a similar number of subjects and channels. 
The work in \cite{touyama2009eeg} explored EEG-based user identification using P300 evoked potentials. The paper achieved a user classification accuracy of up to 97.6\% with 5 subjects using a single EEG channel. 
The authors of \cite{Yu2019} performed a steady-state visual-evoked potentials task involving 8 subjects with 9 EEG channels and achieved a user identification accuracy of 97\% using a modified ConvNet CNN after extracting low-frequency component features from the EEG. 
The work presented in \cite{Ashby2011} included 5 subjects and 14 EEG channels and extracted autoregressive coefficients, power spectral density, spectral power, interhemispheric power difference, and interhemispheric channel linear complexity features. The extracted features were classified with an SVM classifier, achieving an accuracy of 97.69\% without majority voting schemes. Our solution, using a CNN without prior feature extraction, results in a higher classification accuracy than the existing works employing a similar number of subjects and EEG channels.
These results demonstrate the EEG functionality of \textsc{GAPses}, which can be potentially used as a biometric tool to 'unlock' the fast EOG functionalities only to the identified user.

%% file: tables/architecture.tex
\begin{table}[t]
\caption{EpiDeNet architecture~\cite{thorir_biocas}}
\label{tab:EpiDeNet_Table}
\begin{threeparttable}
\centering
\begin{tabular}{lllll}
 & \textit{\textbf{Type}} & \textit{\textbf{\#Filters}} & \multicolumn{1}{c}{\textit{\textbf{Kernel}}} & \textit{\textbf{Output}} \\ \hline
\multirow{2}{*}{$\phi^1$}
 & \multicolumn{1}{|l}{Conv2D} & \multicolumn{1}{l}{$4$} & \multicolumn{1}{l}{($1$, $4$)} & \multirow{1}{*}{($4$,$C$,$T$)} \\
 
 & \multicolumn{1}{|l}{MaxPool} & \multicolumn{1}{l}{} & \multicolumn{1}{l}{(1, 8)} &($4$,$C$,$T//8$) \\ \hline
 \multirow{2}{*}{$\phi^2$} & \multicolumn{1}{|l}{Conv2D} & \multicolumn{1}{l}{$16$} & \multicolumn{1}{l}{($1$, $16$)} & \multirow{1}{*}{($16$,$C$,$T//8$)} \\
 & \multicolumn{1}{|l}{MaxPool} & \multicolumn{1}{l}{} & \multicolumn{1}{l}{(1, 4)} & \multicolumn{1}{l}{($16$,$C$,$T//32$)} \\ \hline
 \multirow{2}{*}{$\phi^3$} & \multicolumn{1}{|l}{Conv2D} & \multicolumn{1}{l}{$16$} & \multicolumn{1}{l}{($1$, $8$)} & \multirow{1}{*}{($16$,$C$,$T//32$)} \\
 & \multicolumn{1}{|l}{MaxPool} & \multicolumn{1}{l}{} & \multicolumn{1}{l}{($1$, $4$)} & \multirow{1}{*}{($16$,$C$,$T//128$)} \\ \hline
\multirow{2}{*}{$\phi^4$} & \multicolumn{1}{|l}{Conv2D} & \multicolumn{1}{l}{$16$} & \multicolumn{1}{l}{($16$, $1$)} & \multirow{1}{*}{($16$,$C$,$T//128$)} \\
 & \multicolumn{1}{|l}{MaxPool} & \multicolumn{1}{l}{} & \multicolumn{1}{l}{($D$, $1$)} & \multirow{1}{*}{($16$,$C//4$,$T//128$)} \\ \hline
 \multirow{2}{*}{$\phi^5$} & \multicolumn{1}{|l}{Conv2D} & \multicolumn{1}{l}{$16$} & \multicolumn{1}{l}{($8$, $1$)} & \multirow{1}{*}{($16$,$C//4$,$T//128$)} \\
 & \multicolumn{1}{|l}{AdaptiveAveragePool} & \multicolumn{1}{l}{} & \multicolumn{1}{l}{} & \multirow{1}{*}{($16$,$1$,$1$)} \\ \hline
\multirow{1}{*}{$\phi^6$} & \multicolumn{1}{|l}{Dense} & \multicolumn{1}{l}{} & \multicolumn{1}{l}{} & \multirow{1}{*}{2} \\
\end{tabular}
\begin{tablenotes}\footnotesize
\item[] $C$ = number of channels, $T$ = number of time samples, $D = 1$ for EOG data and $D = 4$ for EEG data
\vspace{-10pt}
\end{tablenotes}
\end{threeparttable}
\end{table}

%% file: tables/SoA_comparison.tex
\begin{table*}[h!]
\begin{center}
\begin{threeparttable}[b]
\caption{Comparison to \ac{SoA} smart glasses for \ac{EEG} and \ac{EOG} acquisition.}
\label{table:soa_comparison)}
\scriptsize{
\begin{tabular}
{
p{0.5in}>
{\centering\arraybackslash}p{2in}>
{\centering\arraybackslash}p{0.9in}>
{\centering\arraybackslash}p{1.2in}>
{\centering\arraybackslash}p{0.4in}>
{\centering\arraybackslash}p{0.4in}>
{\centering\arraybackslash}p{0.5in}>
{\centering\arraybackslash}p{0.5in}>
{\centering\arraybackslash}p{0.6in}
}

\toprule[0.20em]

\textbf{Device} & \textbf{\# of biopotential channels (Application)} & \textbf{Electrodes} & \textbf{Power consumption [mW]} &\textbf{SoC}  &\textbf{Performance} \newline \textbf{[Mop/s]} & \textbf{Energy efficiency} \newline \textbf{~[Mop/s/mW]}\\
\midrule
AttentivU \cite{Kosmyna2019_attentivU_MIT} & 1 EEG (AW) \newline 1 EOG (visual inspection) & Silver & 111 (stream) & \SI{12}{MHz} MCU & N/A & N/A 
\\
\midrule[0.12em]
\cite{Lee2020_smart_glasses_KU_KIST_Korea} & 1 EEG (AW, SSVEP) \newline 1 EOG (visual inspection, eye movement classification) & Carbon nanotube/polydimethylsiloxane composite & 300 (stream) & ATmega128 & 16 (int8) &  0.24 (int8)
\\
\midrule[0.12em]

e-Glass \cite{Sopic2018_e_glass_epilepsy} & 4 EEG (SD $^{1}$) & N/A & 32.4 (online SD $^{1}$) & STM32L151 & 32 (int32) & 2.35 (int32)
\\
\midrule[0.12em]

\textbf{\textsc{GAPses}} \newline \textbf{(this work)} & \textbf{8 EEG (AW, SSVEP, MM, Brainmetrics)} \newline \textbf{3 EOG (visual inspection, eye movement classification) $^{2}$} & \textbf{Soft conductive rubber with Ag/AgCl coating} & \textbf{30 (stream) \newline 16.28 (online eye movement classification)}  & \textbf{GAP9} & \textbf{3'000 (FP32)} & \textbf{74 (FP32)} \\
\midrule[0.12em]

\end{tabular}}
AW: Alpha Waves, SSVEP: Steady-State Visual Evoked Potential, MM: Motor Movement classification, SD: Seizure Detection, FP: floating point \\ (1) results based on online dataset, (2) concurrent acquisition of up to 8 channels\\

\end{threeparttable}
\end{center}
\vspace{-6mm}
\end{table*}

%% file: Sections/07_Discussion.tex
\section{Comparison to state-of-the-art ExG-based smart glasses}
\label{sect:discussion}

Table~\ref{table:soa_comparison)} summarizes the positioning of \textsc{GAPses} in the context of existing smart glasses for \ac{EEG} and \ac{EOG} acquisition. \textsc{GAPses} stands out with a notably higher channel count, featuring 8 EEG and 3 EOG channels, while existing solutions are either restricted to a single modality or a single channel per modality. In terms of power consumption, the total system power consumption of \textsc{GAPses} when streaming raw data is significantly lower compared to related works (\SI{30}{mW} compared to \SI{111}{mW} in \cite{Kosmyna2019_attentivU_MIT} and \SI{300}{mW} \cite{Lee2020_smart_glasses_KU_KIST_Korea}). Additionally, to the best of our knowledge, \textsc{GAPses} is the only device offering substantial edge processing capabilities, enabling computations at the edge with high energy efficiency.

%% file: Sections/08_Conclusion.tex
\section{Conclusion}\label{sec:conclusion}
In this paper, we presented \textsc{GAPses}, a smart-glasses platform for fully dry and wearable \ac{EEG} and \ac{EOG} measurement. We described the system design, which involved the development of custom ExG electrodes based on soft conductive rubber (to ensure comfort and good signal quality for extended wearing durations), custom mechanical frames, and electronics interfaces to a PULP ExG acquisition and processing platform. 

We first validated \textsc{GAPses'} functionality across various common tasks such as alpha wave detection, \ac{SSVEP}, and \ac{MM} classification.
Additionally, we showcased the device's potential in two end-user applications. Using the device as an EOG-based HMI end-to-end, we demonstrated an accuracy as high as \SI{96.78}{\%} in an eye-movement classification task that discriminates among 11 classes, also demonstrating an \ac{ITR} as high as \SI{161.43}{bit/min}.
Using the device as an \ac{EEG}-based BrainMetrics solution, we demonstrated a subject-specific identification task with average sensitivity and specificity as high as \SI{98.87}{\%} and \SI{99.86}{\%}, respectively. 
These two applications are deployed on the device and consume only \SI{24}{\micro\joule} and \SI{121}{\micro\joule} per inference (for the EOG and EEG application, respectively), operating at an average power during inference of only \SI{16.28}{\milli\watt} and \SI{26.54}{\milli\watt} (for the EOG and EEG application, respectively). 
We envision the use of GAPses in the following manner for the two presented applications: The user puts on the glasses, which recognize the owner using the EEG-based brainmetrics application in a 1 vs all fashion. Once the owner is recognized, the \ac{EOG} \ac{HMI} application is enabled for continuous use. In this mode, the average system power consumption is \SI{12.4}{mW}, allowing for continuous acquisition and online processing of \ac{EOG} signals for over \SI{22}{h} with a small \SI{75}{mAh} battery.

Overall, \textsc{GAPses} sets a new SoA by offering multi-channel acquisition and energy-efficient onboard processing of \ac{EEG} and \ac{EOG} signals in a compact, user-friendly glasses form factor, thus addressing many of the limitations found in previous designs. Future research will focus on leveraging these advancements in unconstrained study settings to expand the device's applications. Additionally, thanks to the ease of use and non-stigmatizing form factor, \textsc{GAPses} also appear as a promising platform for seizure detection relying on a reduced number of temporal channels. In fact, as demonstrated by prior works \cite{eeg-transformer-journal, ingolfsson2024minimizing}, seizure detection can be implemented successfully on-device when using only a small number of temporal channels, compatible with glasses form factors. Finally, future work will also focus on the extensive characterization of a large variety of different artifacts with a large population size, as large-scale artifact analysis is still needed for enabling the adoption of this technology in everyday life conditions.